\documentclass{JHEP3}

\usepackage{amsmath}
\input epsf
\usepackage{epsfig}
\usepackage{amssymb}
\usepackage{graphics}
\usepackage[active]{srcltx}
\usepackage{cancel}

\usepackage{setspace}
\usepackage{array,multirow,bigdelim,arydshln}
\usepackage{shuffle}

\newcommand{\bea}{\begin{eqnarray}}
	\newcommand{\eea}{\end{eqnarray}}
\newcommand{\bean}{\begin{eqnarray*}}
	\newcommand{\eean}{\end{eqnarray*}}
\newcommand{\nn}{\nonumber \\}

\def\W #1{\widetilde{#1}}
\def\WH #1{\widehat{#1}}

\def\d{\partial}

\def\a{{\alpha}}
\def\b{{\beta}}

\def\eref#1{(\ref{#1})}

\def\Label#1{\label{#1}%
	\smash{\hbox to0pt{\raise1ex\hbox{\tiny[#1]}\hss}}}

\def\co{\,,}
\def\ed{\,.}

\title{Generation Function For One-loop Tensor Reduction}
\author{ Bo Feng$^{ab}$\footnote{Emails:  fengbo@csrc.ac.cn} \\
	{$^a$\small Beijing Computational Science Research Center, Beijing 100084, China\\
		$^b$Peng Huanwu Center for Fundamental Theory, Hefei, Anhui, 230026, China}}
\date{\today}
\abstract{For loop integrals, the reduction is the standard method. Having an efficient way to find reduction coefficients is an
	important topic in scattering amplitudes.  In this paper, we present the generation functions of reduction coefficients for general  one-loop integrals with
	arbitrary tensor rank in numerator.
	
}


\newpage
\begin{document}

\section{Motivation}

As the bridge connecting the theoretical frame and the experiment data, scattering amplitude is always one of  central concepts in the quantum field theory. Its efficient computation (including higher order contributions in perturbation) becomes the necessary,
especially with the coming of the LHC experiment \cite{NLOMultilegWorkingGroup:2008bxd}. The on-shell program\footnote{There are many
works in this topic. For introduction, please see following two books \cite{Henn-book,Elvang-book}. Some early works
with on-shell concept are \cite{Bern:1994zx,Bern:1994cg,Cachazo:2004kj,Britto:2004nc,Britto:2004ap,Britto:2005fq}.} in the scattering amplitudes, as the outcome of the challenge, has made the calculations of one-loop amplitude straightforward \cite{Ellis:2011cr}.

In general the loop computation can be divided into two parts: the construction of integrands and then doing the integration.
Although the construction of integrands  using Feynman diagrams are well established, sometimes it is not the economic way to
do so. Looking for a better way to construct integrands is one of current research directions\footnote{For example, the unitarity cut method
proposed in \cite{Bern:1994zx,Bern:1994cg,Britto:2004nc} uses on-shell tree level amplitudes as input. }. For the second part, i.e., doing the integration, a very useful method
is the reduction. It has been shown that any integral can be written as the linear combination of some basis integrals (called the master
integrals) with coefficients as the rational function of external momenta, polarization vectors, masses and spacetime dimension.  Using the idea of reduction,
loop integration can be separated into two parallel tasks: the computation of master integrals and the algorithm to efficiently extract
the reduction coefficients.  Progresses in any one task will enable us to do more and more complicated integrations (for a nice
introduction of recent developments, see \cite{Weinzierl:2022eaz}).

For the reduction, it can be classified into two categories: the reduction at the integrand level and the reduction at the integral level.
The reduction at the integrand level can be systematically solved using the computational algebraic geometry \cite{Ossola:2006us,Mastrolia:2011pr,Badger:2012dp,Zhang:2012ce}. For the reduction at the integral level, the first proposal is
the famous Passarino-Veltman reduction (PV-reduction) method \cite{Passarino:1978jh}. There are other proposals, such as the
Integration-by-Part (IBP) method \cite{Chetyrkin:1981qh,Tkachov:1981wb,Laporta:2000dsw,vonManteuffel:2012np,vonManteuffel:2014ixa,Maierhofer:2017gsa,Smirnov:2019qkx},
the unitarity cut method \cite{Bern:1994zx,Bern:1994cg,Britto:2004nc,Britto:2005ha,Britto:2006sj,Anastasiou:2006jv,Anastasiou:2006gt,Britto:2006fc,
	Britto:2007tt,Britto:2010um},
Intersection number \cite{Mastrolia:2018uzb,Frellesvig:2019uqt,Mizera:2019vvs,Frellesvig:2020qot,Caron-Huot:2021xqj,Caron-Huot:2021iev}. Although there are a lot of
developments for the reduction at the integral level, it is still desirable to improve them by current complexity of computations.

In recent papers \cite{Feng:2021enk,Hu:2021nia,Feng:2022uqp,Feng:2022rfz,Feng:2022iuc} we have introduced the auxiliary vector $R$
to improve the traditional PV-reduction method. Using $R$ we can construct the differential operators and then establish  algebraic recurrence relation to determine reduction coefficients analytically. This method has also been generalized to two-loop sunset
diagram (see \cite{Feng:2022iuc}) where the original PV-reduction method is hard to be used. When using
the auxiliary vector $R$ in the IBP method, the efficiency of reduction has also been improved as shown in  \cite{Chen:2022jux,Chen:2022lue}.

Although the advantage of using auxiliary vector $R$ has been demonstrated from various aspects, the algebraic recursive structure
makes it
still hard to have a general understanding of reduction coefficients for higher and higher tensor ranks in the numerators of
integrands.
Could we get more understanding of the analytical structures of reduction coefficients by this method? As we will show in this paper, indeed we can get more if we probe the reduction problem from
a new angle.
The key idea is the concept of  {\bf generation function}. In fact,  generation function is well known in physics and mathematics. Sometimes the coefficients of a series is hard to imagine, but the series itself is easy to write down. For example, the Hermite Polynomial $H_n(x)$ can be read out from the generation  function
\bea e^{2t x-t^2}=\sum_{n=0}^\infty H_n(x) {t^n\over n!}~\ed~~~~\label{gen-1}\eea
Thus we can ask that if we sum the reduction coefficients of different tensor ranks together, could we get a simpler answer?
For  reduction problem, the numerator of tensor rank $k$ is given by $(2\ell\cdot R)^k$ in our method and we need to see how
to sum them together.  There are many ways to sum them. Two typical ones are
\bea \psi_1(t)=\sum_{n=0}^\infty t^n (2\ell\cdot R)^n={1\over 1-t(2\ell\cdot R)}~\co~~~~~~~\psi_2(t)=\sum_{n=0}^\infty { (2\ell\cdot R)^n t^n\over n!}=e^{t(2\ell\cdot R)}~\ed~~~~\label{gen-2}\eea
In this paper, we will focus on the generation function of the type $\psi_2(t)$ because it is invariant under the differential action, i.e, ${d e^x\over dx}=e^x$. We will see that the generation functions satisfy simple differential equations, which can be solved
analytically.

The plan of the paper is following. In the section two, we present the generation function of  reduction coefficients of tadpole
integral. The tadpole example is a little bit trivial, thus in the section three, we discuss carefully how to find
generation functions for the bubble integral, which is the simplest nontrivial example. With the experience obtained
for bubble, we  present  the solution for general one-loop integrals in the section four. To demonstrate the frame established
in section four, we discuss briefly the triangle example in the section five. Finally, a brief summary and discussion are given in the
section six. Some technical details have been collected in the Appendix, where in the Appendix A, the solution of two typical
differential equations is presented, while the solution of recursive relation for the bubble has been explained in the Appendix B.


\section{Tadpole}

With above brief discussion,   we  start from the simplest case, i.e., the tadpole topology to discuss the generation function. Summing all tensor ranks properly we have\footnote{The mass dimension of parameter $t$ is $-2$.}
\bea I_{tad}(t,R)\equiv \int d\ell { e^{t (2\ell\cdot R)}\over \ell^2-M^2}=c_{1\to 1}(t,R,M) \int d\ell {1\over \ell^2-M^2}~\co~~~\label{Gen-tad-1}\eea
where $c_{1\to 1}(t,R,M)$ is the generation function of reduction coefficients and for simplicity we have defined $
\int d \ell_{i}(\bullet ) \equiv\int {d^D\ell_i\over i\pi^{D/2}}(\bullet )$.
To find closed analytic expression for $c_{1\to 1}(t,R,M)$ we establish the corresponding differential equation.
Acting with $\d_R$ we have
\bea & & {\d \over \d R}\cdot {\d\over \d R} I_{tad}(t,R) =\int d\ell { 4 t^2 \ell^2e^{t (2\ell\cdot R)}\over \ell^2-M^2}
=4 t^2 M^2 I_{tad}(t,R) ~~~~\label{Gen-tad-4}\eea
at one side, and
$\left({\d \over \d R}\cdot {\d\over \d R}c_{1\to 1}(t,R,M) \right)\int d\ell {1\over \ell^2-M^2}$
at another side, thus we get
\bea  {{\d \over \d R}\cdot {\d\over \d R}c_{1\to 1}(t,R,M)= 4 t^2 M^2 c_{1\to 1}(t,R,M)}~\ed~~~\label{Gen-tad-6}\eea
By the Lorentz invariance,  $c_{1\to 1}(t,R,M)$ is the function of $r=R\cdot R$ only, i.e.,
$c_{1\to 1}(t,R,M)=f(r)$. It is easy to see that differential equation \eref{Gen-tad-6} becomes
\bea \boxed{ 4r f''+ { 2D}f'-  {4 t^2M^2}f =0}~\co~~~\label{Gen-tad-14}\eea
which is the form \eref{typ-1-1} studied in the Appendix A.
This second order differential equation has two singular points $r=0$ and $r=\infty$, where  the singular point $r=0$ is canonical. The solution has been given  in \eref{typ-1-32}.
Putting $A=4,B=2D,C=-4t^2 M^2$ in \eref{typ-1-17} and the boundary condition $c_0=1$, we get immediately
\bea c_{1\to 1}(t,R,M)=\sum_{n=0}^\infty { (t^2 M^2 r)^n\over n!  \left({D\over 2} \right)_n}= ~_0 F_1(\emptyset; {D\over 2}; t^2M^2 r)~\ed~~~\label{Gen-tad-sol-1-9}\eea

Before ending this section, let us mention that when we do the reduction for other topologies, we will meet the reduction of $\int d\ell { e^{t (2\ell\cdot R)}\over (\ell-K)^2-M^2}$. Using the momentum shifting, it is easy to see that
\bea \int d\ell { e^{t (2\ell\cdot R)}\over (\ell-K)^2-M^2}=\int d\ell { e^{t (2(\ell+K)\cdot R)}\over \ell^2-M^2}= e^{2t (K\cdot R)}c_{1\to 1}(t,R,M) \int d\ell {1\over \ell^2-M^2}~\ed~~~\label{Gen-tad-sol-1-10}\eea
The results shows the advantage of using generation function with exponential form.

\section{Bubble}

Having found the generation function of tadpole reduction, we move to the first nontrivial example, i.e.,  the generation function of bubble reduction, which is defined through
\bea & & I_{bub}(t,R)\equiv  \int d\ell { e^{t (2\ell\cdot R)}\over (\ell^2-M_0^2)((\ell-K)^2-M_1^2)}\nn
& = & c_{2\to 2} \int d\ell {1\over (\ell^2-M_0^2)((\ell-K)^2-M_1^2)}+ c_{2\to 1;\WH{1}}\int d\ell { 1\over (\ell^2-M_0^2)}+ c_{2\to 1;\WH{0}}\int d\ell { 1\over ((\ell-K)^2-M_1^2)}~\ed~~~\label{Gen-bub-1-1}\eea
For simplicity, we have not written down  variables of reduction coefficients explicitly. If written explicitly, it will be $c(t,R,K;M_0,M_1)$ or
$c(t, R^2, K\cdot R, K^2;M_0,M_1)$ if using the Lorentz contraction form.

The generation form \eref{Gen-bub-1-1} can produce some nontrivial relations among these generation functions of  reduction coefficients easily. Noticing that\footnote{Comparing to the shifting symmetry discussed in \eref{Gen-bub-1-1-2}, one can also consider the symmetry with $\ell\to -\ell$. For this one, we have $R\to -R$ and $K\to -K$. If using the variables $R^2,K\cdot R$, it is invariant. In other words, the symmetry $\ell\to -\ell$ is trivial. }
\bea & & I_{bub}(t,R)\equiv  \int d\ell { e^{t (2\ell\cdot R)}\over (\ell^2-M_0^2)((\ell-K)^2-M_1^2)}=e^{t(2K\cdot R)}\int d\W \ell { e^{t (2\W\ell\cdot R)}\over ((\W\ell +K)^2-M_0^2)(\W \ell^2-M_1^2)} \nn
& = &e^{t(2K\cdot R)}\left\{  c_{2\to 2}(t, R^2, -K\cdot R, K^2;M_1,M_0) \int d\ell {1\over (\ell^2-M_0^2)((\ell-K)^2-M_1^2)}\right.
\nn
& & + c_{2\to 1;\WH{0}}(t, R^2, -K\cdot R, K^2;M_1,M_0) \int d\ell { 1\over (\ell^2-M_0^2)}\nn
& & \left.+ c_{2\to 1;\WH{1}}(t, R^2, -K\cdot R, K^2;M_1,M_0) \int d\ell { 1\over ((\ell-K)^2-M_1^2)}\right\}~\co~~~\label{Gen-bub-1-1-2}\eea
 we have
\bea c_{2\to 2}(t, R^2, K\cdot R, K^2;M_0,M_1) & = & e^{t(2K\cdot R)} c_{2\to 2}(t, R^2, -K\cdot R, K^2;M_1,M_0) ~\co~~~\label{Gen-bub-1-1-2a}   \\
c_{2\to 1;\WH{1}}(t, R^2, K\cdot R, K^2;M_0,M_1) & = &e^{t(2K\cdot R)} c_{2\to 1;\WH{0}}(t, R^2, -K\cdot R, K^2;M_1,M_0)~\co~~~\label{Gen-bub-1-1-2b} \\
c_{2\to 1;\WH{0}}(t, R^2, K\cdot R, K^2;M_0,M_1) & = &e^{t(2K\cdot R)} c_{2\to 1;\WH{1}}(t, R^2, -K\cdot R, K^2;M_1,M_0)~~~~\label{Gen-bub-1-1-2c}\eea
by comparing  \eref{Gen-bub-1-1-2} with \eref{Gen-bub-1-1}.
The first relation \eref{Gen-bub-1-1-2a} can be a consistent check for $c_{2\to 2}$ while the second relation \eref{Gen-bub-1-1-2b} and the third relation \eref{Gen-bub-1-1-2c} tell us that we need to compute only one of $c_{2\to 1;\WH{i}}$ functions. Another useful check is the mass dimension. Since the mass dimension of $t$ is $(-2)$, we have
\bea [c_{2\to 2}]=0,~~~~[c_{2\to 1}]=-2~\ed~\label{Gen-bub-1-1-2d}\eea
%
%

\subsection{Differential equations}

Now we will write down differential equations for
these generation functions. Acting  $\d_R\cdot \d_R$  on both sides of \eref{Gen-bub-1-1} we have
\bea & & \d_R\cdot \d_R I_{bub}(t,R)= \int d\ell { 4t^2 \ell^2e^{t (2\ell\cdot R)}\over (\ell^2-M_0^2)((\ell-K)^2-M_1^2)}\nn
& = & 4 t^2 M_0^2 I_{bub}(t,R)+ 4 t^2 e^{t(2K\cdot R)} \int d\ell { e^{t (2\ell\cdot R)}\over (\ell^2-M_1^2)}~\co~~~\label{Gen-bub-1-2}\eea
thus we derive
\bea  \d_R\cdot \d_R c_{2\to 2}(t, R^2, K\cdot R, K^2;M_0,M_1)&=& 4 t^2 M_0^2 c_{2\to 2}(t, R^2, K\cdot R, K^2;M_0,M_1)~\co~~~\label{Gen-bub-1-2a}  \\
 \d_R\cdot \d_R c_{2\to 1;\WH{1}}(t, R^2, K\cdot R, K^2;M_0,M_1)&= & 4t^2M_0^2 c_{2\to 1;\WH{1}}(t, R^2, K\cdot R, K^2;M_0,M_1) ~\co~~~\label{Gen-bub-1-2b}\\
 \d_R\cdot \d_R c_{2\to 1;\WH{0}}(t, R^2, K\cdot R, K^2;M_0,M_1)&= & 4t^2M_0^2 c_{2\to 1;\WH{0}}(t, R^2, K\cdot R, K^2;M_0,M_1) \nn
& & ~~~+4 t^2 e^{t(2K\cdot R)} c_{1\to 1}(t,R^2,M_1)~\ed~~~\label{Gen-bub-1-2c}\eea
Acting with  $K\cdot \d_R$ we have
\bea & & K\cdot \d_R I_{bub}(t,R)= \int d\ell { t(2K\cdot \ell) e^{t (2\ell\cdot R)}\over (\ell^2-M_0^2)((\ell-K)^2-M_1^2)}=\int d\ell { t(D_0-D_1 +f) e^{t (2\ell\cdot R)}\over (\ell^2-M_0^2)((\ell-K)^2-M_1^2)}\nn
& = & t f I_{bub}(t,R)-t \int d\ell { e^{t (2\ell\cdot R)}\over (\ell^2-M_0^2)}+ t e^{t(2K\cdot R)} \int d\ell { e^{t (2\ell\cdot R)}\over (\ell^2-M_1^2)}~\co~~~\label{Gen-bub-1-3}\eea
where $f=K^2-M_1^2+M_0^2$, thus we  derive
\bea & & K\cdot \d_R c_{2\to 2}(t, R^2, K\cdot R, K^2;M_0,M_1)=t f c_{2\to 2}(t, R^2, K\cdot R, K^2;M_0,M_1)~\co~~~\label{Gen-bub-1-3a}  \\
& & K\cdot \d_R c_{2\to 1;\WH{1}}(t, R^2, K\cdot R, K^2;M_0,M_1)=t f c_{2\to 1;\WH{1}}(t, R^2, K\cdot R, K^2;M_0,M_1)-t c_{1\to 1}(t,R^2,M_0)~\co~~~\label{Gen-bub-1-3b}\\
& & K\cdot \d_R c_{2\to 1;\WH{0}}(t, R^2, K\cdot R, K^2;M_0,M_1)=t f c_{2\to 1;\WH{0}}(t, R^2, K\cdot R, K^2;M_0,M_1)+ t e^{t(2K\cdot R)} c_{1\to 1}(t,R^2,M_1)~\ed~~~~~\label{Gen-bub-1-3c}\eea
 Above two groups of differential equations can be uniformly  written as
\bea   \d_R\cdot \d_R c_T &=&   4t^2M_0^2 c_T+4 t^2 \xi_R h_T ~\co~~\label{Bub-uni-2a} \\
 K\cdot \d_R c_T &= & t f c_T+ t \xi_K h_T ~\co~~\label{Bub-uni-2b}
\eea
where $h_T$ is the possible non-homogenous contribution coming from lower topology (tadpole).
For different type $T$ we have
\bea T=\{2\to 2\}: &~~~& h_T=0 ~~~~~{\rm or}~~~~~\xi_R=\xi_K=0~\co~~\label{Bub-uni-3a} \\
T=\{2\to 1;\WH 1\}: &~~~& h_T=c_{1\to 1}(t,R^2,M_0),~~~\xi_R=0,~~~\xi_K=-1~\co~~\label{Bub-uni-3b} \\
T=\{2\to 1;\WH 0\}: &~~~& h_T=e^{t(2K\cdot R)} c_{1\to 1}(t,R^2,M_1),~~~\xi_R=1,~~~\xi_K=+1~\ed~~\label{Bub-uni-3c}\eea
Generation functions are   functions of $f(r,p)$ with $r=R^2$ and $p=K\cdot R$. It is easy to work out that
\bea  {\d\over \d R^\mu} &= &2\eta_{\rho\mu} R^\rho \d_r + K_\mu \d_p,~~~~K\cdot {\d\over \d R^\mu} =  2p \d_r+ K^2 \d_p  ~\co~~~\label{Gen-bub-1-4b} \\
\eta^{\mu\nu}{\d\over \d R^\nu}{\d\over \d R^\mu}
& = &  (4 r \d^2_r + 4p \d_p\d_r  + K^2 \d^2_p + 2D \d_r)~\co~~~\label{Gen-bub-1-4a}
 \eea
thus \eref{Bub-uni-2a} and \eref{Bub-uni-2b} can be written as
\bea & & (4 r \d^2_r + 4p \d_p\d_r  + K^2 \d^2_p + 2D \d_r-4 t^2 M_0^2) c_T=4 t^2 \xi_R h_T~\co~~\label{Bub-uni-2a-1} \\
& & ( 2p \d_r+ K^2 \d_p-tf) c_T =t \xi_K h_T ~\ed~~\label{Bub-uni-2b-1} \eea
Equations \eref{Bub-uni-2a-1} and \eref{Bub-uni-2b-1} are the differential equations we need to solve. We will present two ways to solve them. One is by the series expansion of the naive variables $r,p$. This is the method used in \cite{Feng:2021enk,Hu:2021nia}. However, as we will show, using the idea of generation function, the powers of $r,p$ are independent to each other, thus the recursion relations
become simpler and
can be solved explicitly.  Another method is to solve the differential equation directly and get more compact and  analytical expression. An important lesson from the second method is that  the right variables to do the series expansion are not $r,p$ but their proper combination.

\subsection{The series expansion}

In this subsection, we will present the solution in the form of series expansion of $r,p$.
Writing\footnote{Since we consider the generation functions, the $n,m$ are free indices, while in previous works \cite{Feng:2021enk,Hu:2021nia} with fixed tensor rank $K$,   $n,m$ are constrained by $2n+m=K$.
One can see that many manipulations are simplified using the idea of generation functions.  }
\bea c=\sum_{n,m=0}^\infty c_{n,m} r^n p^m,~~~~~h_c=\sum_{n,m=0}^\infty h_{n,m} r^n p^m~\co~~\label{Bub-uni-4}\eea
and putting them to \eref{Bub-uni-2a-1} and \eref{Bub-uni-2b-1} we get following equations
\bea 0 & = &  2(n+1)(2n+2m+D) c_{n+1,m}+ K^2(m+2)(m+1) c_{n,m+2}
-4 t^2 M_0^2 c_{n,m}-4t^2 \xi_R h_{n,m} ~~n,m\geq 0~\co~~~~~~~~\label{Bub-uni-5a} \\
0 & = &  2(n+1) c_{n+1,m}+K^2 (m+2) c_{n,m+2}-t f c_{n,m+1}-t \xi_K h_{n,m+1} ~~~~m,n\geq 0~\co~~\label{Bub-uni-5b}  \\
0 & = & K^2  c_{n,1}-t f c_{n,0}-t \xi_K h_{n,0}~~~~~n\geq 0~\ed~~\label{Bub-uni-5c} \eea
Using \eref{Bub-uni-5a} and \eref{Bub-uni-5b} we can solve
\bea c_{n+1,m} & = & { 4 M_0^2 t^2 \over 2(n+1) (D+2n+m-1)} c_{n,m} - { t f (m+1) \over 2(n+1) (D+2n+m-1)} c_{n,m+1}\nn & & +{ 4 t^2 \xi_R h_{n,m}- t \xi_K (m+1) h_{n,m+1} \over 2(n+1) (D+2n+m-1)}~\co~~~~~~~~\label{Bub-uni-6a}\\
c_{n,m+2} & = &  { -4 M_0^2 t^2 \over (m+2) (D+2n+m-1)K^2} c_{n,m} + { t f (D+2n+2m)\over (m+2)(D+2n+m-1) K^2} c_{n,m+1}\nn & & +{ -4 t^2 \xi_R h_{n,m}+t \xi_K (D+2n+2m) h_{n,m+1} \over 2(n+1) (D+2n+m-1)}~\ed~~~~~~~~~\label{Bub-uni-6b}\eea
Now setting $m=0$ in \eref{Bub-uni-6a} and combining \eref{Bub-uni-5c} we can solve
\bea c_{n+1,0} & = &  { (-t^2 f^2+ 4 M_0^2 t^2 K^2) \over 2(n+1) (D+2n-1)K^2} c_{n,0}
-{ t^2( \xi_K f-4 \xi_R K^2) h_{n,0}+ t\xi_K K^2 h_{n,1} \over 2(n+1) (D+2n-1)K^2}~\co~~~~~\label{Bub-uni-7a} \\
c_{n,1} & = & { t f\over K^2} c_{n,0}+ { t \xi_K h_{n,0}\over K^2}~\ed~~~~~\label{Bub-uni-7b}\eea
Using equation \eref{Bub-uni-7a} we can recursively solve all $c_{n,0}$ starting from the boundary condition
$c_{0,0}=1$ for $c_{2\to 2}$ or $c_{0,0}=0$ for $c_{2\to 1}$. Knowing all $c_{n,0}$ we can use
\eref{Bub-uni-7b} to get all $c_{n,1}$. To solve all $c_{n,m}$, we use
 \eref{Bub-uni-5a} and \eref{Bub-uni-5b} again, but now  solve
\bea c_{n,m+1} & = & { 4 M_0^2 t \over f(m+1)} c_{n,m}-{2(n+1)(D+2n+m-1) \over t f (m+1)} c_{n+1,m}+{ 4t \xi_R h_{n,m}-\xi_K (m+1) h_{n,m+1}}~\co~~~~~\label{Bub-uni-8a}\\
c_{n,m+2} & = & {-2 (n+1)(D+2n+2m) \over (m+1)(m+2) K^2}c_{n+1,m}+{4 t^2 M_0^2 \over (m+1)(m+2) K^2}c_{n,m} + {4 t^2\xi_R h_{n,m}\over (m+1)(m+2) K^2}~\ed~~~~~~\label{Bub-uni-8b} \eea
Both equations can be used recursively to solve $c_{n,m}$. After using one of them to get all $c_{n,m}$,  another one becomes nontrivial consistent check.  Among them \eref{Bub-uni-8a} is better,
since it solves $(m+1)$ from $m$.

Using above algorithm, we present  the first few terms of generation functions for comparison.
For $c_{2\to 2}$ we have
\bea c_{2,2} & = & 1+{f t\over K^2} p+{ (D f^2- 4 K^2 M_0^2)t^2\over 2(D-1)(K^2)^2} p^2+{ (4K^2 M_0^2 -f^2) t^2 \over 2(D-1) K^2} r\nn
& & + { f((2+D) f^2-12 K^2 M_0^2) t^3\over 6(D-1) (K^2)^3} p^3- {f(f^2-4 K^2 M_0^2) t^3\over 2(D-1) (K^2)^2} rp +...~~~\label{bub-c-exp-1}\eea
and
\bea c_{2\to 1;\WH 1} & = & 0-{t\over K^2}p- {D f t^2\over 2(D-1)(K^2)^2} p^2+{ f t^2\over 2(D-1)K^2} r\nn
& & -{ ( D(2+D) f^2-8(D-1) K^2 M_0^2) t^3\over 6(D-1)D (K^2)^3} p^3+{ (D f^2-4(D-1) K^2 M_0^2)t^3\over 2(D-1) D (K^2)^2} rp+...~~~\label{bub-c-exp-2}\eea
and
\bea c_{2\to 1;\WH 0} & = & 0 + {t\over K^2}p+{(-D \W f+4(D-1) K^2)t^2\over 2(D-1) (K^2)^2} p^2+{\W f t^2\over 2(D-1) K^2} r+ { (D f\W f+4(D-1) K^2 M_1^2)t^3\over 2 (D-1)D (K^2)^2}rp \nn
&& + { (6D K^2(K^2(D-2)+D M_0^2)-2(D+2)(3D-2)K^2 M_1^2+D(D+2)\W f^2)t^3\over 6(D-1)(K^2)^2}p^3+.. ~~~\label{bub-c-exp-3}\eea
where we have defined $\W f=K^2-M_0^2+M_1^2=-f+2K^2$.

Here we have presented the general recursive algorithm. In the Appendix B, we will show that these recursion relations can be
solved explicitly, i.e., we find  explicit expressions for all coefficients $c_{n,m}$.

\subsection{The analytic solution}

In previous subsection, we have present the solution using the series expansion. In this subsection, we will solve the two differential equations \eref{Bub-uni-2a-1} and \eref{Bub-uni-2b-1} directly.

Let us start from \eref{Bub-uni-2b-1} first. To solve it, we define following new variables
\bea x& = & K^2 r- p^2,~~~~~y=p~\co~~~\label{bub-xy} \eea
then \eref{Bub-uni-2b-1} becomes
\bea & & ( 2p \d_r+ K^2 \d_p-tf) c(r,p) = (K^2 \d_y-t f) c(x,y)= t \xi_K h_T ~\co~~\label{Bub-uni-2b-2} \eea
where we have used
\bea p &= & y,~~~~~~r={x+y^2\over K^2},~~~~
\d_r =K^2 \d_x,~~~
\d_p =   -2 y \d_x+\d_y~\ed~~\label{Bub-uni-2a-4}\eea
The differential equation \eref{Bub-uni-2b-2} can be solved as (see the discussion in the Appendix,
for example, \eref{typ-2-5})
\bea c(x,y)= {1\over K^2} e^{{tf\over K^2}y}\left(G(x) +\int_{0}^y dw  e^{-{tf\over K^2}w} t \xi_K h_T(x,w) \right)~\co~~\label{Bub-uni-2b-5}\eea
where the function $G$ depends only on $x$, while $h_T\equiv h_T(x,y)$ is the function of both $x,y$.

Now we consider the equation \eref{Bub-uni-2a-1}. The first step is to simplify it by writing
\bea
& & (4 r \d^2_r + 4p \d_p\d_r  + K^2 \d^2_p + 2D \d_r)= {1\over K^2}( 2p \d_r+ K^2 \d_p)( 2p \d_r+ K^2 \d_p)+ (4r-{4p^2\over K^2})\d_r^2+2(D-1)\d_r\ed~~~~~~~\label{Bub-uni-2a-3}\eea
Thus using \eref{Bub-uni-2b-1}, \eref{Bub-uni-2a-1} becomes
%
%
%
\bea & &\left( {4 x K^2}\d_x^2+2(D-1)K^2\d_x+ { t^2(f^2-4 K^2 M_0^2)\over K^2} \right) c=\left( -t\xi_K \d_y + { 4 t^2 \xi_R K^2- t^2 \xi_K f\over K^2} \right) h_T\ed~~~\label{Bub-uni-2a-2} \eea
Putting \eref{Bub-uni-2b-5} to \eref{Bub-uni-2a-2} and simplifying we get
\bea & & \left( {4 x K^2}\d_x^2+2(D-1)K^2\d_x+ { t^2(f^2-4 K^2 M_0^2)\over K^2} \right) G(x) \nn
& = & K^2 e^{-{tf\over K^2}y}\left( -t\xi_K \d_y + { 4 t^2 \xi_R K^2- t^2 \xi_K f\over K^2} \right) h_T(x,y)\nn
& & -\left( {4 x K^2}\d_x^2+2(D-1)K^2\d_x+ { t^2(f^2-4 K^2 M_0^2)\over K^2} \right)\int_{0}^y dw  e^{-{tf\over K^2}w} t \xi_K h_T(x,w)   \ed~~~\label{Bub-uni-2a-5} \eea
Equation \eref{Bub-uni-2a-5} is the form of \eref{typ-1-1} which has been discussed in the Appendix. One interesting point is that since the left hand side is independent of $y$, the right hand side should be zero under the action of $\d_y$. One can check that it is indeed true.

Having laid out the frame, we can use it to solve various generation functions.

\subsubsection{The generation function $c_{2\to 2}$}

For this case, we have $h_T=0$, thus using the result \eref{typ-1-17} we can immediately written down
\bea c_{2\to 2}(t,r,p,K^2;M_0,M_1) & = & ~_0F_1(\emptyset;{D-1\over 2};\left({ t^2(4K^2 M_0^2-f^2)x\over 4(K^2)^2}\right)e^{{tf\over K^2}y}|_{x\to K^2 r-p^2, y\to p}~\ed~~\label{Bub-uni-2to2-1} \eea
One can check it with the series expansion \eref{Gen-bub-ser-solve-2-4} given in the Appendix B. Comparing to it, the result \eref{Bub-uni-2to2-1} is very simple and compact. This shows the power of using the generation function. Also, the differential equations \eref{Bub-uni-2a-1} and \eref{Bub-uni-2b-1} tell us the right variables for the series expansion should be $x,y$ instead of the naive variables $r,p$.

\subsubsection{The generation function $c_{2\to 1;\WH 1}$}
For this case we have  $\xi_R=0, \xi_K=-1$ and
\bea h_T(r)= c_{1\to 1}(t,r={x+y^2\over K^2},M_0)=\sum_{n=0}^\infty { (t^2 M_0^2 )^n \left({x+y^2\over K^2} \right)^n\over n!  \left({D\over 2} \right)_n}~\co~~\label{Bub-uni-2to1-1} \eea
which  satisfies the differential equation \eref{Gen-tad-14}. The \eref{Bub-uni-2a-5} becomes
\bea & & \left( {4 x K^2}\d_x^2+2(D-1)K^2\d_x+ { t^2(f^2-4 K^2 M_0^2)\over K^2} \right) G(x) \nn
& = & K^2 e^{-{tf\over K^2}y}t\left(  \d_y + { t  f\over K^2} \right) h_T(x,y)\nn
& & +t\left( {4 x K^2}\d_x^2+2(D-1)K^2\d_x+ { t^2(f^2-4 K^2 M_0^2)\over K^2} \right)\int_{0}^y dw  e^{-{tf\over K^2}w} h_T(x,w) ~\ed~~\label{Bub-uni-2a-6} \eea
A first important check is that the right hand side of \eref{Bub-uni-2a-6} is $y$-independent. Acting ${\d\over \d y}$ at the right hand side we will get
\bea & & t e^{-{tf\over K^2}y}\left\{-tf  \left(  \d_y + { t  f\over K^2} \right) h_T(x,y)
+K^2 \left(  \d^2_y + { t  f\over K^2}\d_y \right) h_T(x,y)\right.\nn
& & \left.+\left( {4 x K^2}\d_x^2+2(D-1)K^2\d_x+ { t^2(f^2-4 K^2 M_0^2)\over K^2} \right)  h_T(x,y)\right\} ~\ed~~\label{Bub-uni-2a-7}\eea
Using
\bea & & \d_x h_T(r)={\d {x+y^2\over K^2}\over \d x}\d_r h_T ={1\over K^2}\d_r h_T,~~~~~
\d_y h_T(r)={\d {x+y^2\over K^2}\over \d y}\d_r h_T={2y\over K^2}\d_r h_T~\co~~\label{Bub-uni-2a-8}\eea
one can check that \eref{Bub-uni-2a-7} is reduced to the differential equation \eref{Gen-tad-14}, thus we have proved the $y$-independent of \eref{Bub-uni-2a-6}.

Setting $y=0$ in \eref{Bub-uni-2a-6} we get
\bea & & \left( {4 x K^2}\d_x^2+2(D-1)K^2\d_x+ { t^2(f^2-4 K^2 M_0^2)\over K^2} \right) G(x) = t^2 f  h_T(x,y=0) ~\co~~\label{Bub-uni-2a-9} \eea
where we have used $\d_y h_T(x,y=0)={2y\over K^2}\d_r h_T=0$. The differential equation \eref{Bub-uni-2a-9} is the form of \eref{typ-1-1} and we get the solution
\bea G(x) & = &{t^2 f\over 4K^2}  G_0(x) \int_0^x dw w^{- {(D-1)\over 2}} G^{-2}_0(x) \int_{0}^w d\xi h_T(\xi,y=0) G_0(\xi) \xi^{ {(D-1)\over 2}-1}~\co~~\label{Bub-2to1-G-1} \eea
where
\bea G_0(x) & = & ~_0F_1(\emptyset; {(D-1)\over 2}; { t^2(4K^2M_0^2-f^2)\over 4(K^2)^2} x)~\ed~~\label{Bub-2to1-G-2} \eea
Putting all together we finally have
\bea c_{2\to 1;\WH 1}(x,y)= {1\over K^2} e^{{tf\over K^2}y}\left(G(x) -t \int_{0}^y dw  e^{-{tf\over K^2}w}   h_T(x,w) \right)~\ed~~\label{Bub-2to1-G-3} \eea

Although we have a very compact expression \eref{Bub-2to1-G-3} for the generation function, in practice it is more desirable to have the series expansion form. In the Appendix A we have introduced three ways.  Here we work out the expansion by direct integration. Using \eref{Bub-uni-2to1-1} we have
\bea & & \int_{0}^y dw  e^{-{tf\over K^2}w}  h_T(x,w)=\sum_{n=0}^\infty { (t^2 M_0^2 )^n \over n!  \left({D\over 2} \right)_n}\int_{0}^y dw  e^{-{tf\over K^2}w}   \left({x+w^2\over K^2} \right)^n~\ed~~\label{bub-c21-exp-1} \eea
To work out the integration, we see that
\bea R(\a)\equiv \int_0^{T} du e^{\a u}={1\over \a} e^{\a u}|_0^{T}={ e^{\a T}-1\over \a}
~\co~~~\label{int-1}\eea
thus
\bea {d^n\over d\a^n} R(\a)&= &\int_0^{T} du e^{\a u} u^n = { (-)^n n!\over \a^{n+1}} \left( e^{\a T} \lfloor e^{-\a T}\rfloor_{\a^{n}}-1\right)\nn & = &{ (-)^n n!\over \a^{n+1}} \left( e^{\a T} \left( e^{-\a T}-\sum_{i=n+1}^\infty {(-\a T)^i\over i!}\right)-1\right) = {- (-)^n n!\over \a^{n+1}}e^{\a T} \sum_{i=n+1}^\infty {(-\a T)^i\over i!}~\co~~~\label{int-2}\eea
where the symbol $\lfloor Y(x)\rfloor_{x^{n-1}}$ means to keep the Taylor expansion up to the order of $x^{(n-1)}$. Using \eref{int-2} we have
\bea & & \int_0^{T} du e^{\a u}( \b+\gamma u^2)^N=\sum_{i=0}^N {N!\over i!(N-i)!} \b^{N-i} \gamma^i \int_0^{T} du e^{\a u} u^{2i}\nn
& = &
\sum_{i=0}^N {N!\over i!(N-i)!} \b^{N-i}\gamma^i{-  (2i)!\over \a^{2i+1}}e^{\a T} \sum_{j=2i+1}^\infty {(-\a T)^j\over j!}~\ed~~~\label{int-3}\eea
Using \eref{int-3} we can evaluate \eref{bub-c21-exp-1} as
\bea & & \int_{0}^y dw  e^{-{tf\over K^2}w}  h_T(x,w)= ye^{-{tf\over K^2} y}\sum_{n=0}^\infty \sum_{j=0}^\infty \sum_{i=0}^n { \left({t^2 M_0^2\over K^2}\right)^n \over  \left({D\over 2} \right)_n}{ (2i)! x^{n-i} y^{2i}\over i!(n-i)! }{\left( {t f y\over K^2} \right)^j\over (j+2i+1)!}~\ed~~\label{bub-c21-exp-2} \eea
The evaluation of $G(x)$ can  be found in \eref{typ-1-12} as
\bea G(x)
 & = &\sum_{n=0}^\infty  \sum_{i=0}^{n-1}  { \left( {D-1\over 2}\right)_{i}\over n! \left( {D-1\over 2}\right)_{n}\left({D\over 2} \right)_i} { f  ( M_0^2 )^i(t^2x)^n (4K^2M_0^2-f^2)^{n-i-1}\over 4^{n-i}(K^2)^{2n-i-1}}~\ed~~\label{bub-c21-exp-3}\eea
Collecting all pieces together, we finally have
\bea c_{2\to 1;\WH 1}(t,r,p,K^2;M_0,M_1) &= &  {1\over K^2} e^{{tf\over K^2}y}\sum_{n=0}^\infty  \sum_{i=0}^{n-1}  { \left( {D-1\over 2}\right)_{i}\over n! \left( {D-1\over 2}\right)_{n}\left({D\over 2} \right)_i} { f  ( M_0^2 )^i(t^2x)^n (4K^2M_0^2-f^2)^{n-i-1}\over 4^{n-i}(K^2)^{2n-i-1}}\nn
& & - {t y\over K^2}\sum_{n=0}^\infty \sum_{j=0}^\infty \sum_{i=0}^n { \left({t^2 M_0^2\over K^2}\right)^n \over  \left({D\over 2} \right)_n}{ (2i)! x^{n-i} y^{2i}\over i!(n-i)! }{\left( {t f y\over K^2} \right)^j\over (j+2i+1)!}~\ed~~\label{bub-c21-exp-4}\eea
This result can be checked with the one given in \eref{2to1-ser-e-3-6}. One can see that  the formula
\eref{bub-c21-exp-4} is much more compact and manifest with various analytic structures.

\section{The general frame}

Having the detail computations in the bubble, in this section we will set up the general frame to
find generation functions for general one-loop integrals with $(n+1)$ propagators. The system has
$n$  external momenta $K_i,i=1,...,n$ and $(n+1)$ masses
$M_j^2,j=0,1,...,n$. Using the auxiliary vector $R$ we have $(n+1)$ {\bf auxiliary scalar products (ASP)} $r=R\cdot R$, $p_i=K_i\cdot R, i=1,...,n$. From the experience in the bubble, we know that these ASP's are not good variables to solve differential equations produced by $\d_R\cdot \d_R$ and $K_i\cdot \d_R, i=1,...,n$. Thus we will discuss how to find these good variables in the first subsection. Then we discuss the differential equations in these new variables in the second subsection and finally their solutions in the third subsection.


\subsection{Finding good variables}

We will denote the good variables by $x$ and $y_i,i=1,...,n$. To simplify the differential equations, we need to impose following conditions
\bea (K_i\cdot \d_R) x=0,~~~~(K_i\cdot \d_R) y_j\sim \delta_{ij},~~~~~\forall i,j=1,...,n~\ed~~\label{tri-var-1}\eea
%
To see there is indeed a solution for \eref{tri-var-1}, let us define the Gram matrix $G$ and the row vector $P^T$ as
\bea G_{ij}=K_i\cdot K_j,~~~~(P^T)_i= K_i\cdot R\ed~~~\label{tri-var-2}\eea
Putting $y_j=\sum_t\b_{j t} p_t$ to  \eref{tri-var-1},  it is easy to see that the condition becomes
\bea K_i\cdot \d_R y_j= \sum_t \b_{jt} (K_t\cdot K_i)\sim \delta_{ij}\co~~~\label{tri-var-3}\eea
thus the matrix $\b$ can be solved as
\bea \b=|G| G^{-1}~\co~~\label{tri-var-4}\eea
where $|G|$ is the Gram determinant. For $x$, let us assume
\bea x=|G| r+ P^T A P,~~~~A^T=A~\ed~~\label{tri-var-5}\eea
Since
\bea K_i\cdot \d_R x &= &  |G| 2p_i+ 2 (P^T)_{R\to K_i} A P~\co~~\label{tri-var-6}\eea
where $P_{R\to K_i}$ means to replace vector $R$ by the vector $K_i$, when collecting all $i$ together,
the right hand side of \eref{tri-var-6} is just $(2|G|I+2 GA) P$, thus we have the solution
\bea A=-|G| G^{-1}\ed~~~\label{tri-var-7}\eea
Putting everything together, we finally have
\bea x & = & |G|(r-  P^T  G^{-1}P),~~~~~~~Y= |G| G^{-1} P~\co~~\label{tri-var-8}  \eea
where the mass dimensions of various quantities are
\bea [|G|]=2n,~~~[(G^{-1})_{ij}]=-2,~~~[A_{ij}]=2(n-1),~~~[x]=2(n+1),~~~[y_i]=2n~\ed~~\label{tri-var-9}\eea
From \eref{tri-var-8} we can solve
\bea
P= {1\over |G|} G Y,~~~~r= {|G|x+Y^T G Y\over |G|^2} \ed~~~\label{tri-var-10}\eea
%

\subsection{The differential equations}

Having found the good variables, we express  differential operators $\d_R\cdot \d_R$ and $K_i\cdot \d_R, i=1,...,n$ using them.
The first step is to use \eref{tri-var-8} to write
\bea {\d\over \d R^\mu} & = & \left( 2|G| R_{\mu}-2|G| {\cal K}^T_\mu  G^{-1} P   \right) \d_x + |G| {\cal K}^T_\mu G^{-1} \d_Y ~\co~~\label{tri-ope-1}\eea
where we have defined ${\cal K}^T=(K_1,...,K_n)$ and $\d_Y^T=(\d_{y_1},...,\d_{y_n})$. Thus we find %
\bea {\cal K}\cdot {\d\over \d R^\mu} =|G|\d_Y,~~~~~~\d_R\cdot\d_R=  2|G|(D-n) \d_x+ 4|G| x \d^2_x+ |G|^2 \d_Y^T  G^{-1}\d_Y~\ed~~\label{tri-ope-2}\eea
The differential equations for  $c_T$, where $T$ denotes different
types of generation functions,  have following pattern
\bea K_i\cdot  \d_R c_T &= & \a_{i} c_T+ H_{T;i},~~i=1,2,...,n~\co\label{tri-ope-3a}\\
\d_R\cdot\d_R c_T &= & \a_R c_T+ H_{T;R} ~\co~~\label{tri-ope-3b} \eea
where $\a_R,\a_i$ are constant (which are independent of $T$) and $H_{T;R},H_{T;i}$  are known functions coming from lower topologies. Using the result \eref{tri-ope-2} and \eref{tri-ope-3a}, we find
\bea
& & |G|^2 \d_Y^T  G^{-1}\d_Y c_T= \a^T_{\cal K} G^{-1}\a_{\cal K}c_T+H^T_{T;\cal K}G^{-1}\a_{\cal K}+ |G| \d_Y^T  G^{-1}H_{T;\cal K}~\co~~\label{tri-ope-3-1} \eea
where $\a^T_{\cal K}=(\a_1,...,\a_n)$ and $H^T_{T;\cal K}=(H_{T;1},...,H_{T;n})$. Thus
the differential equations \eref{tri-ope-3a} can be written as
\bea \left( \d_{y_i}-{\a_i\over |G|}\right) c_T & = & {1\over |G|} H_{T;i}~~i=1,2,...,n~\co\label{tri-ope-4a}\eea
while \eref{tri-ope-3b} becomes
\bea
\left(  4|G| x \d^2_x+ 2|G|(D-n) \d_x+ \W \a_R\right) c_T &= & {\cal H}_{T;R} ~~~\label{tri-ope-4b}   \eea
with
\bea \W \a_R =\a^T_{\cal K} G^{-1}\a_{\cal K}-\a_R,~~~~~~ {\cal H}_{T;R}=-H^T_{T;\cal K}G^{-1}\a_{\cal K}- |G| \d_Y^T  G^{-1}H_{T;\cal K}+ H_{T;R} ~\ed~~\label{tri-ope-4b1} \eea

Having given the differential equations \eref{tri-ope-4a} and \eref{tri-ope-4b}, there is an important point to be mentioned. For  \eref{tri-ope-4a} and \eref{tri-ope-4b} to have solution, functions $H$ are not arbitrary, but must satisfy the {\bf integrability conditions}, which are
\bea & & \left( \d_{y_j}-{\a_j\over |G|}\right) H_{T;i}=\left( \d_{y_i}-{\a_i\over |G|}\right) H_{T;j},~~~\forall i,j=1,2,...,n ~\co~~\label{tri-ope-5-1}\eea
and
\bea \left(  4|G| x \d^2_x+ 2|G|(D-n) \d_x+ \W \a_R\right){1\over |G|}H_{T;i}=\left( \d_{y_i}-{\a_i\over |G|}\right){\cal H}_{T;R},~~~\forall i=1,2,...,n~\ed~~\label{tri-ope-5-2}\eea

Differential equations \eref{tri-ope-4a} and \eref{tri-ope-4b} are the type of \eref{typ-2-1} and \eref{typ-1-1} respectively in the Appendix A, for which the solution has been presented. In next subsection, we will solve them analytically.


\subsection{Analytic solution}

In this part, we will present the necessary steps of solving above differential equations \eref{tri-ope-4a} and \eref{tri-ope-4b}. Let us solve them one by one. For differential equation \eref{tri-ope-4a} with $i=1$, using the result \eref{typ-2-5} in the Appendix A, we have
\bea c_T(x,y) & = & e^{{\a_1\over |G|}y_1}\left(F_T(x,y_2,...,y_n) +{1\over |G|}\int_{0}^{y_1} dw_1  e^{-{\a_1\over |G|}w_1} H_{T;1}(x,w_1,y_2,...,y_n)\right)~\co~~\label{Solve-KR-1-1}\eea
where $F_T(x,y_2,...,y_n)$ does not depend on the variable $y_1$. Now we act with $\left( \d_{y_2}-{\a_2\over |G|}\right)$ at the both sides of \eref{Solve-KR-1-1} to get the differential equation
\bea \left( \d_{y_2}-{\a_2\over |G|}\right)F_T(x,y_2,...,y_n)&= &- {1\over |G|} \int_{0}^{y_1} dw_1 e^{-{\a_1\over |G|}w_1} \left( \d_{y_2}-{\a_2\over |G|}\right)H_{T;1}(x,w_1,y_2,...,y_n)\nn & & +e^{-{\a_1\over |G|}y_1}{1\over |G|} H_{T;2}(x,y_1,y_2,...,y_n)\ed~~~\label{Solve-KR-2-2}\eea
Using \eref{typ-2-5} we find
\bea & &  F_T(x,y_2,...,y_n)=  e^{{\a_2\over |G|}y_2}F_T(x,y_3,...,y_n)+e^{{\a_2\over |G|}y_2}e^{-{\a_1\over |G|}y_1}{1\over |G|}
\int_{0}^{y_2} dw_2 e^{-{\a_2\over |G|}w_2} H_{T;2}(x,y_1,w_2,...,y_n)\nn
& &  -e^{{\a_2\over |G|}y_2} {1\over |G|} \int_{0}^{y_1} dw_1 e^{-{\a_1\over |G|}w_1}\left( e^{-{\a_2\over |G|}y_2} H_{T;1}(x,w_1,y_2,...,y_n)-H_{T;1}(x,w_1,y_2=0,...,y_n)\right)\ed~~~\label{Solve-KR-2-4}\eea
Putting \eref{Solve-KR-2-4} back to  \eref{Solve-KR-1-1} and doing some algebraic manipulations, we get
\bea c_T(x,y)
& = & e^{{\a_1\over |G|}y_1}e^{{\a_2\over |G|}y_2}F_T(x,y_3,...,y_n)\nn
& & +e^{{\a_2\over |G|}y_2}{1\over |G|}
\int_{0}^{y_2} dw_2 e^{-{\a_2\over |G|}w_2} H_{T;2}(x,y_1,w_2,...,y_n) \nn
& &  +e^{{\a_1\over |G|}y_1}e^{{\a_2\over |G|}y_2} {1\over |G|} \int_{0}^{y_1} dw_1 e^{-{\a_1\over |G|}w_1}H_{T;1}(x,w_1,y_2=0,...,y_n)~\ed~~\label{Solve-KR-2-5}\eea
Repeating above procedure with the action $\left( \d_{y_2}-{\a_2\over |G|}\right)$ we can solve $F_T(x,y_3,...,y_n)$ and then find
\bea & & c_T(x,y)
 =  e^{{\a_1\over |G|}y_1}e^{{\a_2\over |G|}y_2}e^{{\a_3\over |G|}y_3}F_T(x,y_4,...,y_n)\nn
 & & + e^{{\a_3\over |G|}y_3} {1\over |G|} \int_0^{y_3} dw_3 e^{-{\a_3\over |G|}w_3}H_{T;3}(x,y_1,y_2,w_3,...,y_n)\nn
 & & +e^{{\a_2\over |G|}y_2} e^{{\a_3\over |G|}y_3} {1\over |G|}
\int_{0}^{y_2} dw_2 e^{-{\a_2\over |G|}w_2}H_{T;2}(x,y_1,w_2,y_3=0,...,y_n)\nn
 & &+ e^{{\a_1\over |G|}y_1}e^{{\a_2\over |G|}y_2}e^{{\a_3\over |G|}y_3}{1\over |G|} \int_{0}^{y_1} dw_1 e^{-{\a_1\over |G|}w_1}H_{T;1}(x,w_1,y_2=0,y_3=0,...,y_n)~\ed~~\label{Solve-KR-3-6} \eea
By checking \eref{Solve-KR-2-5} and \eref{Solve-KR-3-6} we can see that after solving $n$ first order differential equations \eref{tri-ope-4a} we get
\bea & & c_T(x,y_1,...,y_n) =  e^{\sum_{i=1}^n \a_i y_i\over |G|} F(x)+ {\cal H}_{T;K}~~~~~~\label{Solve-KR-4-1}  \eea
with
\bea {\cal H}_{T;K}  =  {1\over |G|}\W\sum_{i=n}^1 e^{\W\sum_{j=n}^i \a_{j} y_{j}\over |G|}
\int_0^{y_{i}} dw_{i}e^{-\a_{i} w_{i}\over |G|} H_{T; i}(x,y_1,...,y_{i-1},w_{i},0,...,0)~\co~~~~~\label{Solve-KR-4-2b} \eea
where for simplicity we have defined the sum $\W\sum_{i=a}^b$ to mean to take the sum over $(a,a-1,a-2,...,b)$ with $a\geq b$.

Before going to solve the only unknown function $F(x)$, let us check that the form \eref{Solve-KR-4-1} does satisfy the differential equations \eref{tri-ope-4a}. When acting with $\left( \d_{y_k}-{\a_k\over |G|}\right)$ on the both sides, it is easy to see that the first term at the right hand side of \eref{Solve-KR-4-1} and the terms in ${\cal H}_{T;K}$ with $i<k$ give zero contributions since they contain only the factor $e^{-\a_k y_k\over |G|}$ depending on $y_k$. For the term $i=k$ in ${\cal H}_{T;K}$, the action gives
\bea
{1\over |G|} e^{\W\sum_{j=n}^{k+1} \a_{j} y_{j}\over |G|}
 H_{T; k}(x,y_1,...,y_{k-1},y_{k},0,...,0)~\ed~~~~~\label{Solve-KR-4-5}  \eea
For the term $i=k+1$ in ${\cal H}_{T;K}$, the action gives
\bea & &{1\over |G|} e^{\W\sum_{j=n}^{k+1} \a_{j} y_{j}\over |G|}
\int_0^{y_{k+1}} dw_{k+1}e^{-\a_{k+1} w_{k+1}\over |G|}\left( \d_{y_k}-{\a_k\over |G|}\right) H_{T; k+1}(x,y_1,...,y_{k},w_{k+1},0,...,0)\nn
& = &{1\over |G|} e^{\W\sum_{j=n}^{k+1} \a_{j} y_{j}\over |G|}
\int_0^{y_{k+1}} dw_{k+1}e^{-\a_{k+1} w_{k+1}\over |G|}\left( \d_{w_{k+1}}-{\a_{k+1}\over |G|}\right) H_{T; k}(x,y_1,...,y_{k},w_{k+1},0,...,0)~\co~~~~~\label{Solve-KR-4-6}  \eea
where in the second line we have used the integrability condition \eref{tri-ope-5-1}. After partial integration we get
\bea  -{1\over |G|}e^{\W\sum_{j=n}^{k+1} \a_{j} y_{j}\over |G|}H_{T; k}(x,y_1,...,y_{k-1},y_{k},0,...,0)+{1\over |G|}
e^{\W\sum_{j=n}^{k+2} \a_{j} y_{j}\over |G|}H_{T; k}(x,y_1,...,y_{k},y_{k+1},0,...,0)\ed~~~~~~\label{Solve-KR-4-7}  \eea
The first term in \eref{Solve-KR-4-7} cancels the term in \eref{Solve-KR-4-5} and we are left with the second term in \eref{Solve-KR-4-7}. Now the pattern is clear. The $i=k+2$ term in ${\cal H}_{T;K}$ will produce two terms after using the integrability condition and partial integration, the first term will cancel the second term in \eref{Solve-KR-4-7}, while the second term will be the form
\bea {1\over |G|}e^{\W\sum_{j=n}^{k+3} \a_{j} y_{j}\over |G|}H_{T; k}(x,y_1,...,y_{k},y_{k+1},y_{k+2},0,...,0)~\ed~~~~~\label{Solve-KR-4-8} \eea
Continuing to the term $i=n$ in ${\cal H}_{T;K}$ we will be left with ${1\over |G|}H_{T; k}(x,y_1,...,y_{n})$, thus we have proved that \eref{Solve-KR-4-1} does satisfy the differential equations \eref{tri-ope-4a}.

Now we consider the differential equation \eref{tri-ope-4b}. Using the form \eref{Solve-KR-4-1}, we derive
\bea \left(  4|G| x \d^2_x+ 2|G|(D-n) \d_x+ \W \a_R\right)F(x) & = & e^{-\sum_{i=1}^n \a_i y_i\over |G|}\left( {\cal H}_{T;R}-\left(  4|G| x \d^2_x+ 2|G|(D-n) \d_x+ \W \a_R\right){\cal H}_{T;K}\right)~\ed~~~~~~~\label{Solve-KR-5-2} \eea
One important point of \eref{Solve-KR-5-2} is that the right hand side must be $y_i$-independent. To check this point, we act $\d_{y_k}$ at the right hand side to give
\bea & & -{\a_k\over |G|}e^{-\sum_{i=1}^n \a_i y_i\over |G|}\left( {\cal H}_{T;R}-\left(  4|G| x \d^2_x+ 2|G|(D-n) \d_x+ \W \a_R\right){\cal H}_{T;K}\right)\nn
&& + e^{-\sum_{i=1}^n \a_i y_i\over |G|}\left( \d_{y_k}{\cal H}_{T;R}-\left(  4|G| x \d^2_x+ 2|G|(D-n) \d_x+ \W \a_R\right)\d_{y_k}{\cal H}_{T;K}\right)\ed~~~~~~~~\label{Solve-KR-5-3} \eea
Since we have proved
\bea \left( \d_{y_k}-{\a_k\over |G|}\right){\cal H}_{T;K}={1\over |G|} H_{T;k}~\co~~~~~~~\label{Solve-KR-5-4} \eea
\eref{Solve-KR-5-3} is simplified to
\bea  -{\a_k\over |G|}e^{-\sum_{i=1}^n \a_i y_i\over |G|} {\cal H}_{T;R}+ e^{-\sum_{i=1}^n \a_i y_i\over |G|}\left( \d_{y_k}{\cal H}_{T;R}-\left(  4|G| x \d^2_x+ 2|G|(D-n) \d_x+ \W \a_R\right) {1\over |G|} H_{T;k}\right)~\ed~~~~~~~\label{Solve-KR-5-5} \eea
Using the integrability condition \eref{tri-ope-5-2} we get
\bea &  & -{\a_k\over |G|}e^{-\sum_{i=1}^n \a_i y_i\over |G|} {\cal H}_{T;R}+ e^{-\sum_{i=1}^n \a_i y_i\over |G|}\left( \d_{y_k}{\cal H}_{T;R}-\left( \d_{y_k}-{\a_k\over |G|}\right){\cal H}_{T;R}\right)=0~\ed~~~~~~~\label{Solve-KR-5-6} \eea
Having checked the $y$-independent, we can take $y_i$ to be any values at the right hand side of
\eref{Solve-KR-5-2}. From the expression \eref{Solve-KR-4-2b} one can see that if we take $y_1=y_2=...=y_n=0$, we have ${\cal H}_{T;K}=0$, thus \eref{Solve-KR-5-2} is simplified to
\bea \left(  4|G| x \d^2_x+ 2|G|(D-n) \d_x+ \W \a_R\right)F(x) & = &  {\cal H}_{T;R}(x,0,0,...,0)\ed~~~~~~~~\label{Solve-KR-5-2a} \eea
Above differential equation is the form of \eref{typ-1-1} in the Appendix A and we can write down the
solution immediately (see \eref{typ-1-32})
\bea  F(x)  =  F_0(x) \left(f_0+\int_0^x dw {w^{- {(D-n)\over 2}}\over  F^2_0(w)} \int_{0}^w d\xi {1\over A} {\cal H}_{T;R}(x,0,0,...,0) F_0(\xi)\xi^{ {(D-n)\over 2}-1}\right)~~~~~~~~\label{Solve-KR-6-1} \eea
where
\bea F_0(x)= ~_0F_1(\emptyset; {(D-n)\over 2}; {-\W\a_R x\over 4|G|})~\ed~~~~~~~\label{Solve-KR-6-2}  \eea
Putting \eref{Solve-KR-6-2} back to \eref{Solve-KR-4-1} we get the final analytic expression of
generation functions
\bea & & c_T(x,y_1,...,y_n) = {\cal H}_{T;K}+ e^{\sum_{i=1}^n \a_i y_i\over |G|}F_0(x) \nn
& & \left(f_0+\int_0^x dw {w^{- {(D-n)\over 2}}\over  F^2_0(w)} \int_{0}^w d\xi {1\over A} {\cal H}_{T;R}(x,0,0,...,0) F_0(\xi)\xi^{ {(D-n)\over 2}-1}\right)~\ed~~~~~\label{Solve-KR-4-1-sol}  \eea

When we consider the generation functions of reduction coefficients of one-loop integrals with $(n+1)$ propagators, there is a special case, where all $H_{T;i}, H_{T;R}$ are zero. For this case, we can write down immediately the generation function
\bea c_{n+1\to n+1}(R,K_1,...,K_n) = ~_0F_1(\emptyset; {(D-n)\over 2}; {-\W\a_R x\over 4|G|})e^{\sum_{i=1}^n \a_i y_i\over |G|}~\ed~~~~\label{Solve-KR-4-1-sol-spe} \eea
%


%
%
%
%

\section{Triangle}

In this part we present another example, i.e., the  triangle, to demonstrate the general frame laid down in previous section.
The seven generation functions are defined by
\bea & & I_{tri}(t,R)\equiv  \int d\ell { e^{t (2\ell\cdot R)}\over (\ell^2-M_0^2)((\ell-K_1)^2-M_1^2)((\ell-K_2)^2-M_2^2)}\equiv \int d\ell { e^{t (2\ell\cdot R)}\over D_0 D_1 D_2}\nn
& = & c_{3\to 3} \int d\ell {1\over D_0 D_1 D_2}+ c_{3\to 2;\WH{i}}\int d\ell { 1\over \prod_{j\neq i,0}^2D_j}+ c_{3\to 1;i}\int d\ell { 1\over D_i}\ed~~~~\label{Gen-tri-1-1}\eea
%
Using the permutation symmetry and the shifting of loop momentum we can find nontrivial relations among these seven
generation functions. The first group of relations is
\bea c_{3\to 3}(t,R;M_0;K_1,M_1;K_2,M_2) & = & c_{3\to 3}(t,R;M_0;K_2,M_2;K_1,M_1) \co\nn
c_{3\to 2;\WH 0}(t,R;M_0;K_1,M_1;K_2,M_2) & = & c_{3\to 2;\WH 0}(t,R;M_0;K_2,M_2;K_1,M_1)\co\nn
c_{3\to 2;\WH 1}(t,R;M_0;K_1,M_1;K_2,M_2) & = & c_{3\to 2;\WH 2}(t,R;M_0;K_2,M_2;K_1,M_1)\co\nn
c_{3\to 2;\WH 2}(t,R;M_0;K_1,M_1;K_2,M_2) & = & c_{3\to 2;\WH 1}(t,R;M_0;K_2,M_2;K_1,M_1)\co\nn
c_{3\to 1; 0}(t,R;M_0;K_1,M_1;K_2,M_2) & = & c_{3\to 1; 0}(t,R;M_0;K_2,M_2;K_1,M_1)\co\nn
c_{3\to 1; 1}(t,R;M_0;K_1,M_1;K_2,M_2) & = & c_{3\to 1; 2}(t,R;M_0;K_2,M_2;K_1,M_1)\co\nn
c_{3\to 1; 2}(t,R;M_0;K_1,M_1;K_2,M_2) & = & c_{3\to 1; 1}(t,R;M_0;K_2,M_2;K_1,M_1)\ed~~~~\label{Gen-tri-1-2}\eea
The second group of relations is
\bea c_{3\to 3}(t,R;M_0;K_1,M_1;K_2,M_2) & = & e^{2K_1\cdot R} c_{3\to 3}(t,R;M_1;-K_1,M_0;K_2-K_1,M_2)  \co  \nn
c_{3\to 2;\WH 0}(t,R;M_0;K_1,M_1;K_2,M_2) & = &  e^{2K_1\cdot R}c_{3\to 2;\WH 1}(t,R;M_1;-K_1,M_0;K_2-K_1,M_2) \co \nn
c_{3\to 2;\WH 1}(t,R;M_0;K_1,M_1;K_2,M_2) & = &e^{2K_1\cdot R} c_{3\to 2;\WH 0}(t,R;M_1;-K_1,M_0;K_2-K_1,M_2)\co \nn
c_{3\to 2;\WH 2}(t,R;M_0;K_1,M_1;K_2,M_2) & = & e^{2K_1\cdot R} c_{3\to 2;\WH 1}(t,R;M_1;-K_1,M_0;K_2-K_1,M_2)\co \nn
c_{3\to 1; 0}(t,R;M_0;K_1,M_1;K_2,M_2) & = &e^{2K_1\cdot R} c_{3\to 1; 1}(t,R;M_1;-K_1,M_0;K_2-K_1,M_2)\co\nn
c_{3\to 1; 1}(t,R;M_0;K_1,M_1;K_2,M_2) & = &e^{2K_1\cdot R} c_{3\to 1; 0}(t,R;M_1;-K_1,M_0;K_2-K_1,M_2)\co\nn
c_{3\to 1; 2}(t,R;M_0;K_1,M_1;K_2,M_2) & = & e^{2K_1\cdot R}c_{3\to 1; 1}(t,R;M_1;-K_1,M_0;K_2-K_1,M_2)\ed~~~~\label{Gen-tri-1-4}\eea
The third group of relations is
\bea c_{3\to 3}(t,R;M_0;K_1,M_1;K_2,M_2) & = & e^{2K_2\cdot R} c_{3\to 3}(t,R;M_2;K_1-K_2,M_1;-K_2,M_0)\co \nn
c_{3\to 2;\WH 0}(t,R;M_0;K_1,M_1;K_2,M_2) & = &  e^{2K_2\cdot R}c_{3\to 2;\WH 2}(t,R;M_2;K_1-K_2,M_1;-K_2,M_0) \co\nn
c_{3\to 2;\WH 1}(t,R;M_0;K_1,M_1;K_2,M_2) & = &e^{2K_2\cdot R} c_{3\to 2;\WH 1}(t,R;M_2;K_1-K_2,M_1;-K_2,M_0)\co \nn
c_{3\to 2;\WH 2}(t,R;M_0;K_1,M_1;K_2,M_2) & = & e^{2K_2\cdot R} c_{3\to 2;\WH 0}(t,R;M_2;K_1-K_2,M_1;-K_2,M_0)\co\nn
c_{3\to 1; 0}(t,R;M_0;K_1,M_1;K_2,M_2) & = &e^{2K_2\cdot R} c_{3\to 1; 2}(t,R;M_2;K_1-K_2,M_1;-K_2,M_0)\co\nn
c_{3\to 1; 1}(t,R;M_0;K_1,M_1;K_2,M_2) & = &e^{2K_2\cdot R} c_{3\to 1; 1}(t,R;M_2;K_1-K_2,M_1;-K_2,M_0)\co\nn
c_{3\to 1; 2}(t,R;M_0;K_1,M_1;K_2,M_2) & = & e^{2K_2\cdot R}c_{3\to 1; 0}(t,R;M_2;K_1-K_2,M_1;-K_2,M_0)\ed~~~~\label{Gen-tri-1-6}\eea
Using these three groups of relations, we just need to compute three generation functions, for example, $c_{3\to 3}$, $c_{3\to 2;\WH 1}$ and $c_{3\to 1;0}$.
The mass dimensions of them are
\bea [t]=-2,~~~~~~[c_{3\to 3}]=0,~~~~[c_{3\to 2;\WH i}]=-2,~~~~~[c_{3\to 1; i}]=-4~\ed~~~\label{Gen-tri-1-7} \eea
%

\subsection{The differential equations}
Since we have given enough details in the section of bubble, here we will be more briefly.
Using $\d_R \cdot \d_R, K_1\cdot \d_R, K_2\cdot \d_R $ operators, we can find
\bea & &  \d_R\cdot \d_R c_{T}(t,R;M_0;K_1,M_1;K_2,M_2)= 4 t^2 M_0^2 c_T(t,R;M_0;K_1,M_1;K_2,M_2)
+4 t^2 \xi_{T} h_{T}~\co~~~\label{Gen-tri-Diff-1-4} \\
& & K_1\cdot \d_R c_{T}(t,R;M_0;K_1,M_1;K_2,M_2)=t f_1 c_T(t,R;M_0;K_1,M_1;K_2,M_2)
-t\W \xi_{T} \W h_{T}+t \xi_{T} h_T~\co~~~\label{Gen-tri-Diff-2-5}\\
& & K_2\cdot \d_R c_{T}(t,R;M_0;K_1,M_1;K_2,M_2)=t f_2 c_T(t,R;M_0;K_1,M_1;K_2,M_2)
-t\WH \xi_{T} \WH h_T+t \xi_{T} h_T~\co~~~\label{Gen-tri-Diff-3-5}\eea
where $T$ is the index for different generation functions and $f_1=K_1^2-M_1^2+M_0^2$, $f_2=K_2^2-M_2^2+M_0^2$. The various constant $\xi,\W \xi,\WH \xi$ are given in the table
\bea
\begin{tabular}{|c|c|c|c|c|c|c|}
  \hline
  $T$ & $\xi_{T}$ & $h_{T}$ & $\W \xi_T$ & $\W h_T$ & $\WH \xi_{T}$ &  $\WH h_T$\\ \hline \hline
  $\{3\to 3\}$ & $0$ & $0$ & $0$ & $0$ &  $0$ & $0$\\ \hline  \hline
  $\{ 3\to 2;\WH 0\}$ & $1$ & $h_{3\to 2;\WH 0}$  & $0$ & $0$ &  $0$ & $0$\\ \hline
   $\{ 3\to 2;\WH 1\}$ & $0$ & $0$ & $1$ & $\W h_{ 3\to 2;\WH 1}$ &  $0$ & $0$\\ \hline
   $\{ 3\to 2;\WH 2\}$ & $0$ & $0$ & $0$ & $0$ &  $1$ & $\WH h_{3\to 2;\WH 2}$\\ \hline \hline
   $\{ 3\to 1; 0\}$ & $0$ & $0$ & $1$ & $\W h_{ 3\to 1;0}$&  $1$ & $\WH h_{3\to 1;0}$\\ \hline
  $\{ 3\to 1; 1\}$ & $1$ & $h_{3\to 1;1}$ & $0$ & $0$ &  $1$ & $\WH h_{3\to 1;1}$\\ \hline
   $\{ 3\to 1; 2\}$ & $1$ & $h_{3\to 1;2}$ & $1$ & $\W h_{3\to 1;2}$ &  $0$ & $0$ \\ \hline
\end{tabular}~~~~\label{Gen-tri-Diff-xi}
\eea
while $h_T$ are given by
\bea h_{3\to 2; \WH 0}& = & e^{2K_1\cdot R} c_{2\to 2}(t,R^2,(K_2-K_1)\cdot R,(K_2-K_1)^2;M_1,M_2)\co\nn
h_{3\to 1; 1}& = &  e^{2K_1\cdot R} c_{2\to 2;\WH 1}(t,R^2,(K_2-K_1)\cdot R,(K_2-K_1)^2;M_1,M_2)\co\nn
h_{3\to 1; 2} & = & e^{2K_1\cdot R} c_{2\to 2;\WH 0}(t,R^2,(K_2-K_1)\cdot R, (K_2-K_1)^2;M_1,M_2)\co~~~~\label{Gen-tri-Diff-1-4-1}\eea
and $\W h_T$ are given by
\bea \W h_{3\to 2;\WH 1} & = & c_{2\to 2}(t,R^2,K_2\cdot R,K_2^2;M_0,M_2)\co\nn
 \W h_{3\to 1; 0} & = & c_{2\to 1;\WH 1}(t,R^2,K_2\cdot R,K_2^2;M_0,M_2)\co\nn
\W h_{3\to 1;2} & = &  c_{2\to 1;\WH 0}(t,R^2,K_2\cdot R,K_2^2;M_0,M_2)\co~~~~\label{Gen-tri-Diff-2-4} \eea
and $\WH h_T$ are given by
\bea \WH h_{3\to 2;\WH 2} & = & c_{2\to 2}(t,R^2,K_1\cdot R,K_1^2;M_0,M_1)\co\nn
 \WH h_{3\to 1; 0} & = & c_{2\to 1;\WH 1}(t,R^2,K_1\cdot R,K_1^2;M_0,M_1)\co\nn
\WH h_{3\to 1;1} & = &   c_{2\to 1;\WH 0}(t,R^2,K_1\cdot R,K_1^2;M_0,M_1)\ed~~~~\label{Gen-tri-Diff-3-4} \eea
The differential equations \eref{Gen-tri-Diff-1-4}, \eref{Gen-tri-Diff-2-5} and  \eref{Gen-tri-Diff-3-5} are indeed the form
\eref{tri-ope-3a} and \eref{tri-ope-3b} in previous section.

For triangle, the natural variables for generation functions are
\bea r=R\cdot R,~~~p_1=K_1\cdot R,~~~~p_2=K_2\cdot R~\ed~~~\label{Gen-tri-Diff-5-1}\eea
However, the good variables for differential equations \eref{Gen-tri-Diff-1-4}, \eref{Gen-tri-Diff-2-5} and \eref{Gen-tri-Diff-3-5} are $x,y_1,y_2$ as
\bea x& = & [K_1^2 K_2^2-(K_1\cdot K_2)^2]r- K_2^2 p_1^2- K_1^2 p_2^2+(2K_1\cdot K_2) p_1 p_2\nn
& = & |G| r- p_1 y_1- p_2 y_2\co\nn
y_2& =& -(K_1\cdot K_2) p_1+K_1^2 p_2,~~~~~y_1=-(K_1\cdot K_2)p_2+K_2^2 p_1~\co~~~\label{Gen-tri-Diff-8-1}\eea
which are defined in
\eref{tri-var-8} with Gram matrix
\bea G(K_1,K_2)\equiv \left( \begin{array}{cc} K_1^2  & K_1\cdot K_2 \\ K_1\cdot K_2 & K_2^2\end{array}\right),~~~~G^{-1}={1\over |G|} \left( \begin{array}{cc} K_2^2  & -K_1\cdot K_2 \\ -K_1\cdot K_2 & K_1^2\end{array}\right)~\ed~~~\label{Gen-tri-Diff-8-3}\eea
Using the new variables, the differential equations are
\bea & & \left( \d_{y_1}-{\a_1\over |G|}\right) c_T  =  {1\over |G|} H_{T;1},~~~~~~~
\left( \d_{y_2}-{\a_2\over |G|}\right) c_T  =  {1\over |G|} H_{T;2} ~\co~~~\label{Gen-tri-Diff-4b-2}  \\
& & \left(4|G| x \d^2_x+2(D-2) |G|\d_x +\W \a_R\right) c_T= {\cal H}_{T;R}~\co~~~\label{Gen-tri-Diff-4a-2} \eea
where
\bea & & \a_1=tf_1,~~~\a_2=t f_2,~~~~\W \a_R={t^2 f_1^2 K_2^2+t^2 f_2^2 K_1^2-2t^2 f_1 f_2 (K_1\cdot K_2)\over |G|}- 4 t^2 M_0^2~\co~~~\label{Gen-tri-Diff-af} \eea
and
\bea
 & & {\cal H}_{T;R}=-\left(K_2^2 \d_{y_1}-(K_1\cdot K_2)\d_{y_2}+ {\a_1 K_2^2-\a_2(K_1\cdot K_2)\over |G|}\right)H_{T;1}\nn & & -\left(K_1^2 \d_{y_2}-(K_1\cdot K_2)\d_{y_1}+ {\a_2 K_1^2-\a_1(K_1\cdot K_2)\over |G|}\right)H_{T;2} + H_{T;R}~\ed~~~~~~\label{Gen-tri-Diff-4a-2H}\eea
The solution is given by
\bea c_T(x,y_1,y_2)
& = & e^{{\a_1\over |G|}y_1}e^{{\a_2\over |G|}y_2}F^{tri}_T(x) +e^{{\a_2\over |G|}y_2}{1\over |G|}
\int_{0}^{y_2} dw_2 e^{-{\a_2\over |G|}w_2} H_{T;2}(x,y_1,w_2) \nn
& &  +e^{{\a_1\over |G|}y_1}e^{{\a_2\over |G|}y_2} {1\over |G|} \int_{0}^{y_1} dw_1 e^{-{\a_1\over |G|}w_1}H_{T;1}(x,w_1,0)~\co~~~~~~\label{Gen-tri-sol-1}\eea
where
\bea F^{tri}_0(x)= ~_0F_1\left(\emptyset; {(D-2)\over 2}; {-\W\a_R x\over 4|G|}\right)~~~~~~~~~\label{Gen-tri-sol-2}  \eea
and
\bea  F^{tri}(x)  =  F^{tri}_0(x) \left(f_0+\int_0^x dw {w^{ {-(D-2)\over 2}} \over (F^{tri}_0(w))^2} \int_{0}^w d\xi {1\over A} {\cal H}_{T;R}(x,0,...,0) F^{tri}_0(\xi)\xi^{ {(D-2)\over 4}}\right)~\ed~~~~~~~~~\label{Gen-tri-sol-3}\eea
%



{\bf The generation function $c_{3\to 3}$:}
For this one, from the Table \eref{Gen-tri-Diff-xi}, we see that $H_{T;1}=H_{T;2}=H_{T;R}=0$, thus we write down immediately
\bea c_{3\to 3}= e^{{\a_1\over |G|}y_1}e^{{\a_2\over |G|}y_2}  ~_0F_1\left(\emptyset; {(D-2)\over 2}; {-\W\a_R x\over 4|G|}\right)~\ed~~~~~~~~~\label{Gen-tri-sol-3to3}\eea
%

{\bf The generation function $c_{3\to 2}$:}
There are three generation functions $c_{3\to 2;\WH i}$. We want to choose one of them with the simplest $H_{1},H_2,H_R$. Checking with the Table \eref{Gen-tri-Diff-xi}, we see that if we consider $c_{3\to 2;\WH 1}$, we will have $H_R=0, H_2=0$ and
\bea H_{3\to 2;\WH 1;1}(x,y_1,y_2)& = & - c_{2\to 2}(t,R^2,K_2\cdot R,K_2^2;M_0,M_2)\nn
& = & - ~_0F_1\left(\emptyset;{D-1\over 2};\left({ t^2(4K_2^2 M_0^2-f_2^2)(K_2^2r-p_2^2)\over 4(K_2^2)^2}\right)\right)e^{{tf_2\over K_2^2}p_2}~\co~~~~~~~~~\label{Gen-tri-sol-3to2-1}\eea
where we have used the result \eref{Bub-uni-2to2-1} and  expressed $r,p_1,p_2$
using \eref{tri-var-10}
\bea p_1={ K_1^2 y_1+(K_1\cdot K_2) y_2\over |G|},~~~~p_2={ K_2^2 y_2+(K_1\cdot K_2) y_1\over |G|},~~~r={|G|x+ K_1^2 y_1^2+ K_2^2 y_2^2+2(K_1\cdot K_2) y_1 y_2\over |G|^2}\ed~~~~~~~~~\label{Gen-tri-sol-3to2-2}\eea
Thus by \eref{Gen-tri-Diff-4a-2H}  we can find
\bea
& & {\cal H}_{3\to 2;\WH 1;R}= \left( {\a_1 K_2^2-\a_2(K_1\cdot K_2)\over |G|}\right)e^{{tf_2\over K_2^2}p_2} ~_0F_1\left(\emptyset;{D-1\over 2};\left({ t^2(4K_2^2 M_0^2-f_2^2)(K_2^2r-p_2^2)\over 4(K_2^2)^2}\right)\right)\nn
& & +{ t^2(4K_2^2 M_0^2-f_2^2)\over 4(K_2^2)^2} { 2K_2^2 y_1\over |G|}e^{{tf_2\over K_2^2}p_2} ~_0F_1^{(1)}\left(\emptyset;{D-1\over 2};\left({ t^2(4K_2^2 M_0^2-f_2^2)(K_2^2r-p_2^2)\over 4(K_2^2)^2}\right)\right)~\co\eea
where we have defined
\bea ~_A F_B^{(n)} (a_1,...,a_A; b_1,...,b_B;x)= {d^n \over dx^n} ~_A F_B (a_1,...,a_A; b_1,...,b_B;x)~\ed\eea
Putting them back to \eref{Gen-tri-sol-1} we can find the analytic expression. Using it we can  get the explicit series expansion as discussed in the Appendix A.

For generation functions $c_{3\to 1;i}$, we can do the similar calculations. The key is to find $H_{T;i}$ and $H_{T;R}$, which is the
generation functions of one order lower topology. Thus we see the recursive structures of generation functions from lower topologies to
higher topologies. The logic is clear although working out details takes some effects.

\section{Conclusion}

In this paper, we have introduced the concept of generation function for reduction coefficients of loop integrals.
For one-loop integrals, using the recent proposal of auxiliary vector $R$, we can construct two types of differential
operators ${\d \over \d R}\cdot {\d \over \d R}$ and $K_i\cdot {\d \over \d R}$. Using these operators, we can establish
corresponding differential equations for generation functions. By proper changing of variables, these differential
equations can be written into the decoupled form, thus one can solve them one by one analytically. Obviously, one could try to apply the
same idea to discuss the reduction problem for two and higher loop integrals. But with the appearance of irreducible scalar products,
the problem  becomes harder. One can try to use the IBP relation in the Baikov representation \cite{Baikov:1996iu} as did in \cite{Chen:2022jux,Chen:2022lue}.

\section*{Acknowledgments}
We would like to thank Jiaqi Chen, Tingfei Li and Rui Yu for useful discussion and early participant. This work is supported by  Chinese NSF funding under Grant No.11935013, No.11947301, No.12047502 (Peng Huanwu Center).
%

\appendix

\section{Solving differential equations}

As shown in the paper, differential equations for generation functions can be reduced to two typical types.
In this appendix, we present  details of solving these typical  differential equations.

~\\ {\bf The first order differential equation:} The first typical differential equation is following first order differential equation
\bea \boxed{\left( A {d\over dx}+B\right) F(x)= H(x)}~\co~~~\label{typ-2-1}\eea
where $A,B$ are independent of $x$ and $H(x)$ is the known function of $x$.
To solve it, first we solve the homogenous part
\bea \left( A {d\over dx}+B\right) F_0(x)= 0,~~~~~\Longrightarrow~~~ F_0(x) = e^{-{B\over A}x}\ed~~~~\label{typ-2-2}\eea
Then we write $F(x)=F_0(x) F_1(x)$ in \eref{typ-2-1} to get the differential equation for $F_1(x)$ as
\bea A{d\over dx} F_1(x) = F_0^{-1} H(x)~\co~~~\label{typ-2-3}\eea
thus we have
\bea F_1(x)=F_1(x=0)+ \int_{0}^x dt {1\over A} e^{{B\over A}t} H(t)~\ed~~~\label{typ-2-4}\eea
Knowing the special solution of $F(x)$ in \eref{typ-2-4}, the general solution of \eref{typ-2-1} is given by
\bea \boxed{F(x)= F_0(x) F_1(x) + \W \a F_0(x)= e^{-{B\over A}x}\left(\a +\int_{0}^x dw {1\over A} e^{{B\over A}w} H(w)\right)}~\co~~~\label{typ-2-5}\eea
where the constant $\a$ is determined by the boundary condition.

If we want to have the series expansion of $x$, there are several ways to do. The first one is to carry out the integration
\bea \int_{0}^x dt {1\over A} e^{{B\over A}t} H(t)={x\over A} \sum_{n=0}^\infty \sum_{a=0}^\infty h_n x^{n} {1\over (n+a+1)}  {\left({Bx\over A}\right)^{a}\over  a!}~\co~~~\label{typ-2-6}\eea
where we have used the expansion of  $H(x)=\sum_{n=0}^\infty h_n x^n$. The second  way is to put $F(x)=\sum_{n=0}^\infty f_n x^n$ directly to \eref{typ-2-1} to arrive the recursion relation
\bea f_{n+1}={  -B\over (n+1) A}f_n+{h_n\over (n+1) A}\equiv \gamma(n) f_{n}+\rho(n),~~~~n\geq 0~\ed~~~\label{typ-2-7}\eea
The recursive relation \eref{typ-2-7} can be solved as
\bea f_{n+1} &= & f_{0}\prod_{i=0}^n \gamma(i)+\sum_{i=0}^n \rho(i) \prod_{j=i+1}^{n}\gamma(j)~\ed~~~\label{typ-2-8}\eea
It is easy to compute
\bea \Xi[n+1]& = & \prod_{i=0}^{n} \gamma(i)= { (-B)^{n+1}\over (n+1)! A^{n+1}}~~~~~\label{typ-2-9}  \eea
and
\bea & & \rho(i) \prod_{j=i+1}^{n-1}\gamma(j)={ h_i\over (i+1)A}\prod_{j=i+1}^{n-1}\gamma(j)
= {h_i\over -B}\prod_{j=i}^{n-1}\gamma(j)= {h_i\over -B} {\Xi[n]\over \Xi[i]}=
{h_i\over -B} { i!(-B)^{n-i}\over n! A^{n-i}}~\co~~~~\label{typ-2-11} \eea
thus we have
\bea F(x)& = & \sum_{n=0} \left( f_0{ (-B)^{n}\over n! A^{n}}+\sum_{i=0}^{n-1} {h_i\over A} { i!(-B)^{n-1-i}\over n! A^{n-1-i}}\right) x^n~\ed~~~~\label{typ-2-12} \eea
The third way is to use analytic expression \eref{typ-2-5}. We need to compute ${d^n F(x)\over dx^n}$ and then set $x=0$. One can see that, for example,
\bea {d F(x)\over dx} &= &-{B\over A}e^{-{B\over A}x}\left(\a +\int_{0}^x dw {1\over A} e^{{B\over A}w} H(w)\right)+ {1\over A}  H(x)~\co~~~\label{typ-2-13}\eea
thus
\bea {d F(x)\over dx}|_{x=0} &= &-{B\over A}\a + {1\over A}  H(x=0)~\ed~~~\label{typ-2-14}\eea
The important point is that when setting $x=0$ at the end of differentiation,  the integration $\int_0^x dw...=0$, thus we have got rid of integration and all we need to do are the
differentiations over $e^{{B\over A}x}$ and  $H(x)$.

~\\ {\bf The second order differential equation:} The second typical differential equation met in this paper is the  following second order
differential equation
%
\bea \boxed{\left( A x {d^2\over dx^2}+ B {d\over dx}+C\right) F(x)= H(x)}~\co~~~~\label{typ-1-1}\eea
where $A,B,C$ are independent of $x$ and  $H(x)$ is a known function of $x$. Let us to solve it using the series expansion. Writing
\bea F(x)=\sum_{n=0}^\infty f_n x^n,~~~~~~~H(x)=\sum_{n=0}^\infty h_n x^n~\co~~~~\label{typ-1-2}\eea
and putting back to \eref{typ-1-1} we have
\bea & &\sum_{n=0}^\infty h_n x^n=\sum_{n=0}^\infty ( A n(n+1) f_{n+1} x^{n}+ B (n+1) f_{n+1} x^{n}+ C f_n x^n)~\co~~~~\label{typ-1-3}\eea
thus we have the relation
\bea  h_n= (n+1)(B+ A n)  f_{n+1}+ C f_n,~~~n\geq 0~\ed~~~~\label{typ-1-4}\eea
Using it, we can solve\footnote{An important point is that although \eref{typ-1-1} is second order differential equations, but because the $x$ in front of ${d^2\over dx^2}$, around $x=0$ it is essentially the first order differential equation. This explains why using only $f_0$ and known $H(x)$ we can determine $F(x)$ using \eref{typ-1-5}.}
\bea f_{n+1} & = & { h_n\over (n+1)(B+ A n)}- C {1\over (n+1)(B+ A n)} f_n\equiv \gamma(n) f_{n}+\rho(n),~~~~n\geq 0~\ed~~~~\label{typ-1-5}\eea
The recursive relation \eref{typ-1-5} can be solved as
\bea f_{n+1} &= & f_{0}\prod_{i=0}^n \gamma(i)+\sum_{i=0}^n \rho(i) \prod_{j=i+1}^{n}\gamma(j)~\ed~~~~\label{typ-1-8}\eea
It is easy to compute
\bea \Xi[n+1]& = & \prod_{i=0}^{n} \gamma(i)= { (-C)^{n+1}\over (n+1)! A^{n+1}\prod_{i=0}^{n} (B/A+i)}= { (-C)^{n+1}\over (n+1)! A^{n+1}\left( {B\over A}\right)_{n+1}}~\co~~~~\label{typ-1-10}  \eea
where we have used the  {\bf Pochhammer symbol} to simplify the expression\footnote{From the definition one can see that $(x)_{n=0}=1,\forall x$. }
\bea (x)_n= {\Gamma(x+n)\over \Gamma(x)}=\prod_{i=1}^n (x+(i-1))~\ed~~~\label{Poch-1}\eea
Using
\bea & & \rho(i) \prod_{j=i+1}^{n-1}\gamma(j)={ h_i\over (i+1)(B+ A i)}\prod_{j=i+1}^{n-1}\gamma(j)
= {-h_i\over C}\prod_{j=i}^{n-1}\gamma(j)= {-h_i\over C} {\Xi[n]\over \Xi[i]}~~~~~\label{typ-1-11} \eea
 we have
\bea F(x) & = & \sum_{n=0}^\infty f_n x^n= \sum_{n=0}^\infty  x^n\Xi[n]\left\{ f_{0}+\sum_{i=0}^{n-1} {-h_i\over C \Xi[i]}\right\}~\ed~~~~~\label{typ-1-12}\eea
Let us define the following function
\bea H_{A,B,C}(x)=\sum_{i=0}^{\infty} {-h_i x^i\over C \Xi[i]}= \sum_{i=0}^{\infty} {i! A^i\over (-C)^{i+1}} \left( {B\over A}\right)_{i}h_i x^i~\co~~~~~\label{typ-1-13}\eea
which can be considered as the "dual function" of $H(x)$ corresponding to the differential equation \eref{typ-1-1}, then we can write
\bea \sum_{i=0}^{n-1} {-h_i\over C \Xi[i]}=\lfloor H_{A,B,C}(x=1)\rfloor_{x^{n-1}}~\co~~~~~\label{typ-1-14}\eea
where the symbol $\lfloor Y(x)\rfloor_{x^{n-1}}$ means to keep the Taylor series up to the order of $x^{(n-1)}$. Thus $F(x)$ can be written compactly as
\bea F(x) & = & \sum_{n=0}^\infty f_n x^n= \sum_{n=0}^\infty  x^n\Xi[n]\left\{ f_{0}+\lfloor H_{A,B,C}(x=1)\rfloor_{x^{n-1}}\right\}~\ed~~~~~\label{typ-1-15}\eea
For the special case $H(x)=0$, it is easy to see that
\bea F(x)= f_0 F_0,~~~~~~F_0(x)\equiv \sum_{n=0}^\infty { (-C)^{n} x^n\over n! A^{n}\left( {B\over A}\right)_{n}}~\ed~~~~~\label{typ-1-16}\eea
The expression \eref{typ-1-16} is nothing, but the  special case of {\bf generalized hypergeometric function} (see (C.2) of \cite{Weinzierl:2022eaz}), which is defined as
\bea ~_A F_B (a_1,...,a_A; b_1,...,b_B;x)=\sum_{n=0}^\infty { (a_1)_n ... (a_A)_n \over (b_1)_n ...(b_B)_n} {x^n\over n!}~\co~~~\label{Gen-tad-sol-1-8}\eea
thus we have
\bea F(x)= f_0 F_0(x),~~~~~~F_0=~_0F_1(\emptyset; {B\over A}; {-Cx\over A}) ~\ed~~~~~\label{typ-1-17}\eea

Solution in \eref{typ-1-12} is given in the series expansion. We can also write it in the analytic
expression. Writing $F(x)=F_0(x) F_1(x)$ we can find the differential equation of $F_1(x)$ as
\bea H(x) =  \left( F_0(x) A x {d^2\over dx^2}+ (B F_0(x)+2 Ax {d F_0(x)\over dx}) {d\over dx}\right) F_1(x)~\co~~~~~\label{typ-1-21}\eea
which is the first order differential equation of ${d F_1(x)\over dx}=U(x)$.  Using the similar
method as for the differential equation \eref{typ-2-1} we can solve
\bea U(x)={x^{-B\over A}\over F_0^2(x)}\left( \alpha_1+{1\over A}\int_0^x dw H(w) F_0(w) w^{{B\over A}-1}\right) \eea
thus we have 
\bea F(x) & = & F_0(x) \left(\alpha_2+ \int_0^x dt{t^{-B\over A}\over F_0^2(t)}\left( \alpha_1+{1\over A}\int_0^t dw H(w) F_0(w) w^{{B\over A}-1}\right)\right)~\co~~~~~\label{typ-1-28}\eea
where the $\a_1,\a_2$ can be determined using the initial condition of $F(x=0)$ and ${dF(x=0)\over dx}$. Using the expansion of $F(x)$ we see that $\a_2=f_0$ and $\a_1=0$ thus we have
\bea \boxed{ F(x)  =  F_0(x) \left(f_0+\int_0^x dt{t^{-B\over A}\over F_0^2(t)}{1\over A}\int_0^t dw H(w) F_0(w) w^{{B\over A}-1}\right)}~\ed~~~~~\label{typ-1-32}\eea

From \eref{typ-1-32} we can easily obtain the series expansion. For simplicity, let us consider the case $f_0=0$. Using the series expansion 
\bea H(x)=\sum_i h_i x^i,~~~~~F_0(x)=\sum_i f_i x^i\eea
we get 
\bea F_0^2(x) & = & \sum_i x^i \sum_{a+b=i} f_a f_b,~~~
F_0^{-1}(x) =  \sum_i x^i \W f_i,~~~~ \sum_{a+b=s} f_a \W f_b=\delta_{0,s}\eea
Now we have 
\bea & & \int_0^t dw H(w) F_0(w) w^{{B\over A}-1}=\int_0^t dw  w^{{B\over A}-1} \sum_s w^s \sum_{a+b=s} h_a f_b \nn
& = & \int_0^t dw   \sum_s w^{s+{B\over A}-1} \sum_{a+b=s} h_a f_b = \sum_s t^{s+{B\over A}}  \sum_{a+b=s} {1\over a+b+{B\over A}}h_a f_b\eea
and then 
\bea & & \int_0^x dt{t^{-B\over A}\over F_0^2(t)}{1\over A}\int_0^t dw H(w) F_0(w) w^{{B\over A}-1}\nn
&= & \sum_{s=0}^\infty x^{s+1}  \sum_{a+b+c+d=s} {1\over (a+b+c+d)+1}{1\over a+b+{B\over A}} h_a f_b \W f_c\W f_d \eea
Finally we have
\bea & & F_0(x)\int_0^x dt{t^{-B\over A}\over F_0^2(t)}{1\over A}\int_0^t dw H(w) F_0(w) w^{{B\over A}-1}\nn
& = & \left( \sum_i x^i f_i\right) \left(\sum_{s=0}^\infty x^{s+1}  \sum_{a+b+c+d=s} {1\over (a+b+c+d)+1}{1\over a+b+{B\over A}} h_a f_b \W f_c\W f_d \right)\nn
& = & \sum_{s=0}^\infty x^{s+1} \sum_{a+b+c+d+e=s} {1\over (a+b+c+d)+1}{1\over a+b+{B\over A}} h_a f_b \W f_c\W f_d  f_e~~~~~\label{typ-1-35}
\eea
%

\section{The explicit solutions of $c_{n,m}$ for bubble reduction}
In this part, we will show how to get explicit solutions for the recursion relations \eref{Bub-uni-7a}, \eref{Bub-uni-7b}, \eref{Bub-uni-8a} and \eref{Bub-uni-8b}.

\subsection{The generation function  $c_{2\to 2}$}

For this case, we have $h_T=0$, i.e., all $h_{n,m}=0$  in \eref{Bub-uni-7a}, \eref{Bub-uni-7b}, \eref{Bub-uni-8a} and \eref{Bub-uni-8b}. Using \eref{Bub-uni-7a} it is easy to find
\bea c_{N,0} & = & \prod_{n=1}^{N}{ (-f^2+ 4 K^2 M_0^2) t^2 \over 2 K^2 n(D+2n-3)}=\prod_{n=0}^{N-1}{K^2( \b  - \a^2  ) \over 2(n+1) (D+2n-1)}~\co~~~\label{Gen-bub-ser-1-4a-2a}\eea
where we have used the initial condition $c_{0,0}=1$ and defined
\bea \a={t f\over K^2},~~~~\b={4t^2 M_0^2\over K^2}\ed~~~~\label{alpha-beta}\eea
Using \eref{Bub-uni-8a} to compute first few $c_{n,m}$, one can see the pattern
\bea c_{N,m}= {1\over m!}d_{N,m} c_{N,0}~\co~~~\label{Gen-bub-ser-solve-2-4} \eea
where $d_{N,m}$ depends on both $N,m$ and first few $d_{N,m}$ are
\bea d_{N,0}=1,~d_{N,1}=\a,~~~d_{N,2}={ (D+2N) \a^2  -\b \over  (D+2N-1)}\ed~~~\label{afew-dNm}\eea
Putting the form \eref{Gen-bub-ser-solve-2-4} to \eref{Bub-uni-8a} we get the recursion relation
\bea d_{N,m+1}
& = &  { \b\over \a } d_{N,m} - {(\b-\a^2) (D+2N+m-1)\over \a (D+2N-1)} d_{N+1,m}~\ed~~~\label{Gen-bub-ser-solve-2-4b} \eea
Using the formally defined  operator $\WH P$ such that
\bea \WH P f(N,m)= f(N+1,m)~\co\eea
the solution of \eref{Gen-bub-ser-solve-2-4b} can be formally given by
\bea d_{N,M}
& = & \prod_{m=1}^{M}\left( { \b\over \a }  - {(\b-\a^2) (D+2N+m-2)\over \a (D+2N-1)} \WH P\right),~~~M\geq 1,\nn
& := & \left( { \b\over \a }  - {(\b-\a^2) (D+2N+M-2)\over \a (D+2N-1)} \WH P\right)\left( { \b\over \a }  - {(\b-\a^2) (D+2N+(M-1)-2)\over \a (D+2N-1)} \WH P\right)\nn%
& & \times ......\left( { \b\over \a }  - {(\b-\a^2) (D+2N+2-2)\over \a (D+2N-1)} \WH P\right)\left( { \b\over \a }  - {(\b-\a^2) (D+2N+1-2)\over \a (D+2N-1)} \WH P\right)d_{N,0}~\co~~~\label{Gen-bub-ser-solve-2-4f}\eea
where since the appearance of the operator $\WH P$, the ordering of the multiple factors is given explicitly. With a little computation, one can see that
\bea d_{N,M} & = & \sum_{i=0}^M { (-)^i\b^{M-i}(\b-\a^2)^i\over \a^M } d_{N,M;i}~\co~~~\label{Gen-bub-ser-solve-2-4g}\eea
where
\bea d_{N,M;i} & = & \sum_{1;[t_1,...,t_i]}^{M}~\prod_{s=1}^i { D+2(N+s-1)+(M+1-t_s)-2\over D+2(N+s-1)-1},~~~i\geq 1;~~d_{N,M;0}=1~~~~\label{Gen-bub-ser-solve-2-4i}\eea
and the summation sign is defined as
\bea \sum_{1\leq t_1<t_2...<t_i\leq M}:=\sum_{1;[t_1,...,t_i]}^{M}~\ed~~~\label{Gen-bub-ser-solve-2-4h}\eea
Combining \eref{Gen-bub-ser-solve-2-4} with \eref{Gen-bub-ser-1-4a-2a} and \eref{Gen-bub-ser-solve-2-4g}, we have the explicit solution for the generation function of $c_{2\to 2}(t,r,p,K^2;M_0,M_1)$.

\subsubsection{The generation function $c_{2\to 1;\WH 1}$}

For this case, we have $ h_T= c_{1\to 1}(t,r,M_0)$. Using \eref{Gen-tad-sol-1-9} to \eref{Bub-uni-7a} one can  write down
\bea c_{n+1,0}= \gamma(n) c_{n,0}+\rho(n)~~~\label{2to1-ser-6-4-11} \eea
with
\bea \gamma(n)= { K^2(\b-\a^2)\over 2(n+1)(D+2n-1)},~~~~\rho(n)= {t^{2n+1}(M_0^2)^{n} \a \over 2(D+2n-1) (n+1)! \left({D\over 2} \right)_n} ~\co~~\label{2to1-ser-6-4-2} \eea
thus we can solve
\bea c_{n+1,0} &= & c_{0,0}\prod_{i=0}^n \gamma(i)+\sum_{i=0}^n \rho(i) \prod_{j=i+1}^{n}\gamma(j)~\co~~\label{2to1-ser-6-4-3} \eea
where for current case, the initial condition is $c_{0,0}=0$. To find $c_{n,m}$ we write
\bea c_{N,m}& = & {1\over m!} d_{N,m} c_{N,0}- {1\over m!}{t \b^N (K^2)^{N-1}\over  N! 4^N\left({D\over 2} \right)_N} b_{N,m}~~~\label{2to1-ser-e-3-6} \eea
with the first few $d_{N,m}$ and $b_{N,m}$
\bea d_{N,0} & =& 1,~~~~~~~b_{N,0}=0~\co \nn
d_{N,1} & =& \a,~~~~~~~b_{N,1}=1~\co\nn
d_{N,2} & = & { -\b+\a^2(D+2N)\over (D+2N-1)},~~~~~~~~~b_{N,2}={\a (D+2N)\over (D+2N-1) }~\ed~~\label{2to1-ser-e-3-7} \eea
Using the form \eref{2to1-ser-e-3-6} to \eref{Bub-uni-8a} we get the recursion relations
\bea d_{N,m+1} & = & {\b\over \a} d_{N,m}-{(D+2N+m-1) (\b-\a^2)\over \a (D+2N-1)} d_{N+1,m}~\co~~\label{2to1-ser-e-4-3a}\\
b_{N,m+1}& = & {\b\over \a}b_{N,m}-{ \b (D+2N+m-1)\over \a (D+2N)} b_{N+1,m}+{ (D+2N+m-1)\over (D+2N-1)} d_{N+1,m}~\ed~~\label{2to1-ser-e-4-3b}  \eea
The solution of $d_{N,m}$ can be similarly solved using the operator language as in previous subsubsection and we find
\bea d_{N,M} & = & \sum_{i=0}^M { (-)^i\b^{M-i}(\b-\a^2)^i\over \a^M } d_{N,M;i}~~~\label{2to1-ser-e-5-2}\eea
with
\bea d_{N,M;i} & = & \sum_{1;[t_1,...,t_i]}^{M}~\prod_{s=1}^i { D+2(N+s-1)+(M+1-t_s)-2\over D+2(N+s-1)-1},~~~i\geq 1;~~d_{N,M;0}=1~\ed~~\label{2to1-ser-e-5-3}\eea
The solution for $b_{N,m}$ is a little bit complicated because the third term at the right hand side of \eref{2to1-ser-e-4-3b}. To solve it, we write
\bea b_{N,m+1}& = & \W\gamma_N(m)b_{N,m}+\W \rho_N(m) ~\co~~\label{2to1-ser-e-6-1}\eea
where
\bea \W\gamma_N(m)= \left({\b\over \a}-{ \b (D+2N+m-1)\over \a (D+2N)} \WH P\right),~~~~\W \rho_N(m)={ (D+2N+m-1)\over (D+2N-1)} d_{N+1,m} ~\ed~~\label{2to1-ser-e-6-2}\eea
Iterating \eref{2to1-ser-e-6-1} with proper ordering we have
\bea b_{N,m+1} & = & \W\gamma_N(m)\W\gamma_N(m-1)...\W\gamma_N(0) b_{N,0}+\W\gamma_N(m)\W\gamma_N(m-1)...\W\gamma_N(1)
\W\rho_N(0) \nn
& &  +\W\gamma_N(m)\W\gamma_N(m-1)...\W\gamma_N(2)
\W\rho_N(1)+...+\W\gamma_N(m)\W \rho_N(m-1)+\W \rho_N(m) ~\ed~~\label{2to1-ser-e-6-4}\eea
Using
\bea (\WH P)^a \W\rho_N(m)={ (D+2(N+a)+m-1)\over (D+2(N+a)-1)} d_{N+1+a,m} ~~~\label{2to1-ser-e-6-5}\eea
and
\bea & & \W\gamma_N(m)\W\gamma_N(m-1)...\W\gamma_N(m-k)\W \rho_N(m-k-1)\nn
&= &\sum_{i=0}^{k+1} (-)^i\left({\b\over\a}\right)^{k+1}
\left(\sum_{1;[t_1,...,t_i]}^{k+1} \prod_{s=1}^i {D+2(N+s-1)+(m+1-t_s)-1\over D+2(N+s-1)}\right)\nn && \times  { (D+2(N+i)+(m-k-1)-1)\over (D+2(N+i)-1)} d_{N+1+i,m-k-1}~\co~~\label{2to1-ser-e-6-7}
\eea
we finally reach
\bea b_{N,m+1}
& = & { (D+2N+m-1)\over (D+2N-1)} d_{N+1,m}+\sum_{k=0}^{m-1}\left({\b\over\a}\right)^{k+1} \sum_{i=0}^{k+1} (-)^i\nn
& & \times{(D+2(N+i)+(m-k-1)-1)\over (D+2(N+i)-1)} d_{N+1+i,m-k-1}\nn
& &  \times\left(\sum_{1;[t_1,...,t_i]}^{k+1} \prod_{s=1}^i {D+2(N+s-1)+(m+1-t_s)-1\over D+2(N+s-1)}\right)
~\ed~~~~~~~~\label{2to1-ser-e-6-8} \eea
where the condition $b_{N,0}=0$ has been used. The formula \eref{2to1-ser-e-3-6} plus \eref{2to1-ser-6-4-3}, \eref{2to1-ser-e-5-2} and \eref{2to1-ser-e-6-8} gives the explicit
solution for generation function $c_{2\to 1;\WH 1}$.

Knowing $c_{2\to 1;\WH 1}$, we can use \eref{Gen-bub-1-1-2c} to get the generation function $c_{2\to 1;\WH 0}$ or directly compute it using \eref{Bub-uni-7a}, \eref{Bub-uni-7b}, \eref{Bub-uni-8a} and \eref{Bub-uni-8b}.

\subsubsection{The proof of one useful relation}

When we use the improved PV-reduction method with auxiliary vector $R$ to discuss the
reduction of sunset topology, an important reduction relation between different tensor
ranks has been observed  in \cite{Feng:2022iuc}. Later this relation has been studied in \cite{Feng:2022rfz,Chen:2022jux,Linew}. For bubble it is given explicitly by
\bea I_2^{(r)} & = & { (D+2r-4)f p \over (D+r-3) K^2} I_2^{(r-1)}- { (r-1) ( 4M_0^2 p^2+(f^2-4 M_0^2 K^2)r) \over (D+r-3) K^2}I_2^{(r-2)}\nn
& & + {p\over K^2}I_{2;\WH 0}^{(r-1)} + { (r-1) (  (K^2-M_0^2 +M_1^2)r-2 p^2) \over (D+r-3) K^2}I_{2;\WH 0}^{(r-2)} + {-p\over K^2}I_{2;\WH 1}^{(r-1)}+ { (r-1)  f r \over (D+r-3) K^2}I_{2;\WH 1}^{(r-2)}
~\co~~~\label{bub-rank-rec}\eea
where
\bea I_2^{(r)}=\int d\ell {(2\ell\cdot R)^r\over (\ell^2-M_0^2)((\ell-K)^2-M_1^2)},~~~I_{2;\WH 1}^{(r)}=\int d\ell {(2\ell\cdot R)^r\over (\ell^2-M_0^2)},~~~ I_{2;\WH 0}^{(r)}=\int d\ell {(2\ell\cdot R)^r\over (\ell-K)^2-M_1^2}~\ed~~~\label{bub-rank-rec-1}\eea

We can use the series form to check the relation \eref{bub-rank-rec}. Let us define
\bea F[N+2]& = & c^{(N+2)} -\left( {(D+2N) f p c^{(N+1)}-(N+1) ( 4M_0^2 p^2+(f^2-4M_0^2 K^2)r) c^{(N)}\over (D+N-1)K^2}\right)~\co~~\label{Bub-uni-1}\eea
where
\bea c^{(N)}={N!\over t^N} \sum_{n=0}^{[N/2]} c_{n,N-2n} r^n p^{N-2n} ~~~\label{Bub-uni-even-2}\eea
and $c$ can be $c_{2\to 2}$ or $c_{2\to 1}$. Depending on if $N$ is even or odd, the computation details have some differences, thus we consider $N$ even only, while $N$ odd will be similar.
Expanding \eref{Bub-uni-1} we have
\bea F[2N+2] & = & {(2N+2)!\over t^{2N+2}}r^0 p^{2N+2}\left( c_{0,2N+2} -\left( {(D+4N) t f  c_{0,2N+1}\over (2N+2)(D+2N-1)K^2}  -{  4M_0^2 t^2 c_{0,2N}\over 2(N+1) (D+2N-1)K^2}  \right)\right)\nn
& & +{(2N+2)!\over t^{2N+2}} r^{N+1} p^0\left( c_{N+1,0}-\left( -{t^2 (f^2-4M_0^2 K^2) \over (2N+2)(D+2N-1)K^2}  c_{N,0}\right)\right) \nn
& & +{(2N+2)!\over t^{2N+2}}\sum_{n=1}^N r^n p^{2N+2-2n}\left( c_{n,2N+2-2n} -\left({(D+4N)t f \over (2N+2)(D+2N-1)K^2}  c_{n,2N+1-2n}\right.\right.\nn & & \left.\left.-{  4M_0^2t^2 \over (2N+2)(D+2N-1)K^2}  c_{n,2N-2n}  -{t^2(f^2-4M_0^2 K^2) \over(2N+2) (D+2N-1)K^2}  c_{n-1,2N+2-2n}\right)\right)~\ed~~\label{Bub-uni-even-3}\eea
For the term with $r^0 p^{2N+2}$, using the relation \eref{Bub-uni-6b} with $n=0,m=2N$, we find the part inside the bracket is simplified to
\bea { -4 \xi_R t^2 h_{0,2N}+\xi_K t (D+4N) h_{0,2N+1}\over 2 (N+1)(D+2N-1) K^2}~\ed~~\label{Bub-uni-even-4a}\eea
For the term with $r^{N+1} p^{0}$, using the relation \eref{Bub-uni-7a} with $n=N$, we find the part inside the bracket is simplified to
\bea -{ t^2(\xi_K f- \xi_R 4 K^2) h_{N,0}+ \xi_K t K^2 h_{N,1} \over 2(N+1) (D+2N-1) K^2} ~\ed~~\label{Bub-uni-even-4b}\eea
For the term with $r^{n} p^{2N+2-2n}$, the computation is a little bit complicated. First we use \eref{Bub-uni-6a} with $n\to n-1$ and $m\to 2N-2n+2,2N-2n+1,2N-2n$ to write all $c_{i,j}$ with $i=n-1$. Then we use \eref{Bub-uni-6b} with $n\to n-1$ and $m\to 2N-2n+1, 2N-2n$. After doing above two steps and making algebraic simplification, we get
\bea & & { t^3( \xi_K M_0^2- \xi_R f)\over (N+1) (D+2N-1) n K^2} h_{n-1, 2N-2n+1}
+{ t^2( -\xi_K n f+ \xi_R 4K^2(N+1))\over 2(N+1) (D+2N-1) n K^2} h_{n-1, 2N-2n+2}\nn
&  & + { -\xi_K (2N-2n+3) t\over 2 n (D+2N-1)}h_{n-1, 2N-2n+3}~\ed~~\label{Bub-uni-even-4c} \eea
For different $c$ we use the different known $h_T$ as given in \eref{Bub-uni-3a},   \eref{Bub-uni-3b} and \eref{Bub-uni-3c}, thus one can see the relation \eref{bub-rank-rec}
is satisfied.


\end{document}